# First principles investigation of anionic redox in bisulfate lithium battery cathodes

*Pawan Kumar Jha,† Shashwat Singh,† Mayank Shrivastava,$ Prabeer Barpanda† and Gopalakrishnan Sai Gautam*‡*

† Faraday Materials Laboratory (FaMaL), Materials Research Centre, Indian Institute of Science, Bangalore 560012, India

$ Department of Electronic Systems Engineering, Indian Institute of Science, Bangalore 560012, India

‡ Department of Materials Engineering, Indian Institute of Science, Bangalore 560012, India

**Corresponding Author**

* **E-mail:** saigautamg@iisc.ac.in

## Abstract

The search for an alternative high-voltage polyanionic cathode material for Li-ion batteries is vital to improve the energy densities beyond the state-of-the-art, where sulfate frameworks form an important class of high-voltage cathode materials due to the strong inductive effect of the $S^{6+}$ ion. Here, we have investigated the mechanism of cationic and/or anionic redox in $Li_xM(SO_4)_2$ frameworks (M = Mn, Fe, Co, and Ni and $0 \leq x \leq 2$) using density functional calculations. Specifically, we have used a combination of Hubbard *U* corrected strongly constrained and appropriately normed (SCAN+*U*) and generalized gradient approximation (GGA+*U*) functionals to explore the thermodynamic (polymorph stability), electrochemical (intercalation voltage), geometric (bond lengths), and electronic (band gaps, magnetic moments, charge populations, etc.) properties of the bisulfate frameworks considered. Importantly, we find that the anionic (cationic) redox process is dominant throughout delithiation in the Ni (Mn) bisulfate, as verified using our calculated projected density of states, bond lengths, and on-site magnetic moments. On the other hand, in Fe and Co bisulfates, cationic redox dominates the initial delithiation ($1 \leq x \leq 2$), while anionic redox dominates subsequent delithiation ($0 \leq x \leq 2$). In addition, evaluation of the crystal overlap Hamilton population reveals insignificant bonding between oxidizing O atoms throughout the delithiation process in the Ni bisulfate, indicating robust battery performance that is resistant to irreversible oxygen evolution. Finally, we observe both GGA+*U* and SCAN+*U* predictions are in qualitative agreement for the various properties predicted. Our work should open new avenues for exploring lattice oxygen redox in novel high voltage polyanionic cathodes, especially using the SCAN+*U* functional.

## 1 Introduction

Li-ion battery (LIB) technology has become ubiquitous in the 21$^{st}$ century propelling a range of technologies from small-scale portable electronics to large-scale electric vehicles. Since its commercialization by SONY® (circa 1991), the ongoing quest to 'build better batteries' has led to the development of various high energy density electrode materials, electrolytes that are stable over wide voltage windows, and critical advances in cell design. Cathodes form the cornerstone of modern LIBs, which have been conventionally based on 3*d* transition metal (TM) redox (i.e., cationic redox) chemistry.[1] In principle, the capacity and energy density of conventional cathodes are capped by the maximum possible number of electrons exchanged by the TM. In addition to cationic redox, last decade has seen numerous reports on the lattice oxygen acting as an active redox centre leading to capacity improvement,[2-7] which has ushered a new dimension in Li-ion batteries, namely the anionic redox systems.

Reversible electrochemical anionic redox activity has been verified using various sophisticated analytical tools as well as *ab-initio* calculations. The origin of oxygen redox, upon removal of $Li^+$ in Li-rich TM oxides (TMOs), has been attributed to the formation of peroxo-like $(O_2)^{n-}$ species using the 'reductive coupling' mechanism.[5, 8] An important thermodynamic driving force for the activation of oxygen redox is the increase in energy of non-bonded O 2*p* states, which can occur in Li-rich TMOs, resulting in the facile extraction of electrons from O states rather than TM states.[4] Till date, the study of anionic redox are mostly focused on (Li-rich and disordered) oxide compounds.[7, 9] Here, we use density functional theory (DFT[10, 11])-based calculations to explore potential anionic redox in a bisulfate class of polyanionic Li-cathodes, using the insights provided by Li-rich TMOs.

Polyanionic intercalation frameworks exhibit rich diversity in chemistry, polymorphism, tuneable redox potential as provided by the inductive effect, stability in terms of structure and thermodynamics. Various compounds with polyanionic units [$(XO_4)_m^{n-}$: X = B, P, S, Si, Ti, V, Mo, W, As etc.] have been explored as LIB cathodes over the last few decades.[12, 13] Among them, $SO_4$-based polyanionic systems can exhibit the highest redox potentials given the high oxidation state of +6 on the S atoms resulting in a strong inductive effect that reduces the electron density around $O^{2-}$ ions.[12, 14] For example, sulfate systems such as lithium iron bisulfate [$Li_2Fe(SO_4)_2$] and sodium iron polysulfate [$Na_2Fe_2(SO_2)_3$] display $Fe^{3+}/Fe^{2+}$ average redox voltages of 3.83 V (vs. $Li^+/Li$) and 3.8 V (vs. $Na^+/Na$) respectively.[15, 16] Notably, bisulfate $Li_2Fe(SO_4)_2$ displays the highest redox potential, so far, reported for any fluorine free iron-based cathode.

Non-iron TM bisulfates, i.e., $Li_2M(SO_4)_2$ (M = Mn, Co, Ni and Zn), are found to stabilize in monoclinic marinite (m-$P2_1/c$) or orthorhombic (o-$Pbca$) phase depending on the synthesis conditions (except for the Mn bisulfate which exists only in the monoclinic phase).[17-21] Typically, mechanical strain, as induced by energy ball milling, stabilize the metastable (and denser) orthorhombic polymorph. The enthalpy of formation of bisulfates decreases (i.e., becomes more negative) with the increase in the ionic radius of the TM in both polymorphs (except for the Ni analogues).[18, 22] The bisulfates can deliver high cationic redox voltage approaching 3.8 V, as indicated by the example of $Li_2Fe(SO_4)_2$.[16] In addition, they can trigger anionic redox activity at high voltage (> 5 V) leading to high capacity, which is difficult to experimentally realize due to the lack of high-voltage electrolytes.[22]

The current work examines the origin and extent of anionic redox activity in bisulfate $Li_2M(SO_4)_2$ family of polyanionic compounds, where M = Mn, Fe, Co, and Ni. The Hubbard *U* corrected strongly constrained and appropriately normed (SCAN+*U*) exchange correlation (XC) functional within the framework of DFT is employed to explore the underlying redox mechanisms.[23,24, 25] To the best of our knowledge, this is the first study employing SCAN+*U* XC functional to examine polyanionic intercalation materials. In addition to SCAN+*U*, we also used the Hubbard *U* corrected generalized gradient approximation (GGA+*U*) XC functional to determine the phase stability, electronic structure, charge transfer mechanism, and corresponding topotactic average Li-intercalation voltage.[26, 27] We demonstrate that a higher degree of covalent bonding between the M and O atoms results in higher voltage. We also observe that the SCAN+*U* XC functional gives a better qualitative description of polyanionic systems, consistent with the lower degree of self-interaction errors in SCAN compared to GGA.[25, 28] Finally, we examine the underlying mechanism of oxygen lattice redox by calculating and analysing our partial density of states (pDOS), crystal overlap Hamilton population (COHP), charge density difference isosurfaces, variations in M-O bond lengths, and on-site magnetic moment data. We hope that our study will open new avenues for using the SCAN+*U* functional and anionic redox in candidate polyanionic cathodes to develop higher energy density LIBs.

## 2 Crystal Structure

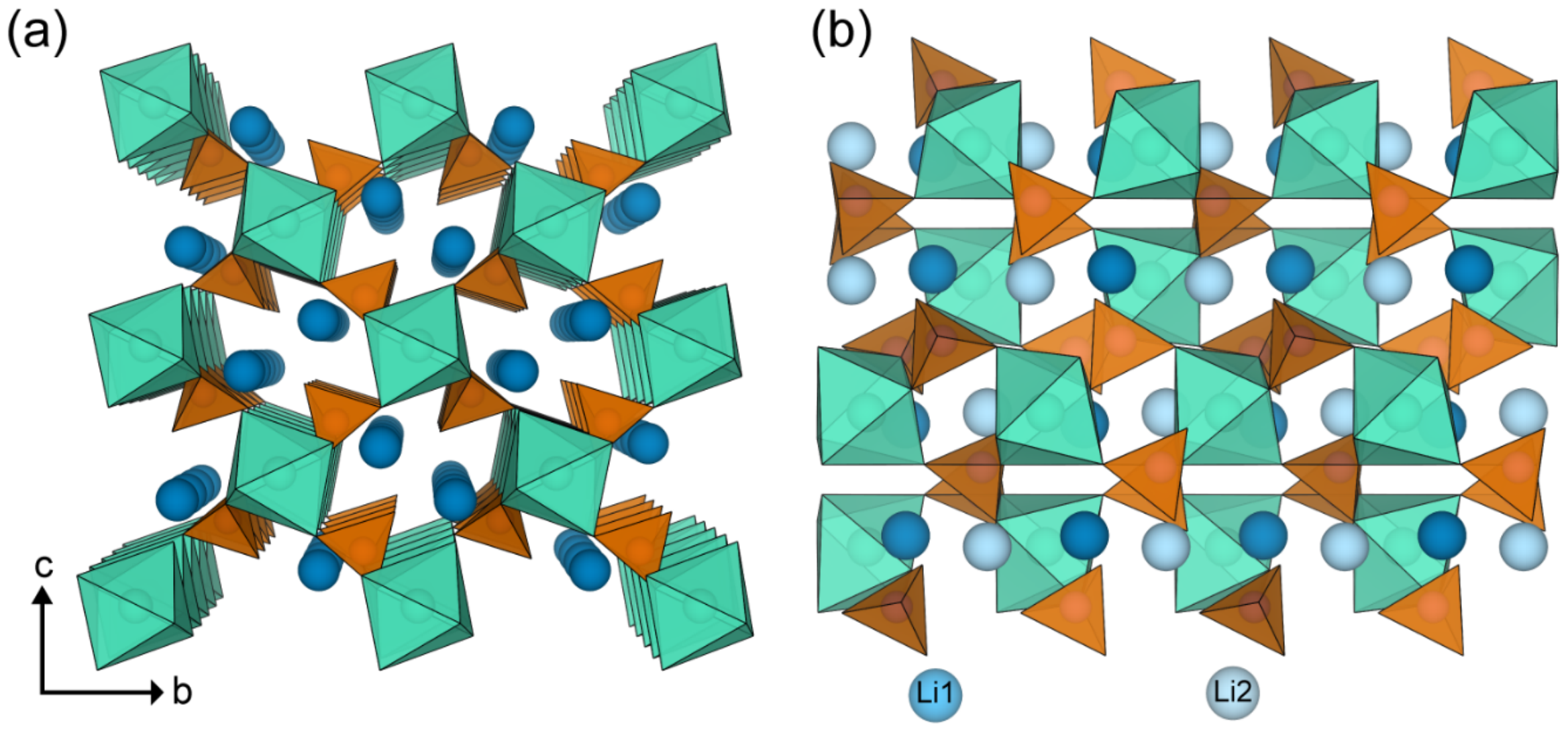

**Figure 1**: Crystal structure of (a) monoclinic and (b) orthorhombic polymorph of $Li_2M(SO_4)_2$, shown along the *a* axis. Li atoms, $MO_6$ octahedra, and $SO_4$ tetrahedra are presented in blue, cyan, and brown, respectively. Distinct Li sites in the orthorhombic polymorph are presented as dark (Li1) and light (Li2) blue color.

The crystal structure of the monoclinic marinite m-$Li_2M(SO_4)_2$ (M = Mn, Fe, and Co), as displayed in **Figure 1a**, has been reported by Tarascon and co-workers,[16, 19, 21] while some of us have recently synthesized the Ni-analogue, m-$Li_2Ni(SO_4)_2$.[22] The marinite framework is built with $MO_6$ octahedra (green polyhedra in **Figure 1a**) and $SO_4$ tetrahedra (brown polyhedra). All corners of $MO_6$ octahedra are linked with six $SO_4$ tetrahedra, resulting in a star-like pattern in the *a-c* plane (i.e., along *b* axis, not shown in **Figure 1a**). Conversely, out of four, only three corners of $SO_4$ tetrahedra are connected to $MO_6$ octahedra with the fourth vertex pointing to a running "tunnel" along *a* axis, where Li atoms reside (blue spheres). $LiO_6$ octahedra share edges with $MO_6$ octahedra, other $LiO_6$ octahedra, and $SO_4$ tetrahedra, apart from sharing their vertices with $SO_4$ tetrahedra and $MO_6$ octahedra as well.

The orthorhombic framework, o-$Li_2M(SO_4)_2$ (M = Co and Fe), as shown in **Figure 1b**, was modelled by Tarascon and others, using a prior report of o-$Li_2Ni(SO_4)_2$ material.[29] The orthorhombic polymorph also consists of $MO_6$ octahedra connected through corner-sharing of a $SO_4$ tetrahedra, similar to the marinite polymorph. However, the arrangement of $MO_6$ octahedra in the orthorhombic framework is denser than the monoclinic and hence breaks the tunnel of Li-sites along the *a* axis, resulting in two distinct Li sites, namely Li1 (dark blue spheres in **Figure 1b**), and Li2 (light blue spheres). Similar to the monoclinic structure, one corner of $SO_4$ tetrahedra is not shared with $MO_6$ octahedra and is instead shared with both Li1 and Li2 sites. While Li1 sites are octahedrally coordinated (forming $LiO_6$), Li2 sites coordinate with five oxygen neighbours forming $LiO_5$ units. The combination of Li1 and Li2 sites results in edge-sharing “zig-zag” chains of Li sites along the *b* axis (view shown in **Figure S1**).

### 3 Computational Methods

DFT calculations were performed using the Vienna *ab initio* simulation package (VASP) using projected augmented wave (PAW) potentials, where the one electron wave functions were expanded up to a maximum kinetic energy of 520 eV using a plane-wave basis.[30-32] Γ-centred Monkhorst-Pack $k$-point meshes, with a density of 48 $k$-points per Å, were used to sample the Brillouin zone.[33] We allowed the cell volume, shape, and positions of all ions to relax for all structures, with the total energies and atomic forces converged to within $10^{-6}$ eV and |0.01| eV/Å, respectively.[33, 34] All calculations were spin polarized and we initialized the magnetic moments of all TM ions with a ferromagnetic ordering. In case of SCAN+$U$ calculations, we used $U$ values of 2.7, 3.1, 3.0, and 2.5 eV for Mn, Fe, Co, and Ni respectively, as derived in previous work.[25, 28] The $U$ values in the case of GGA+$U$ calculations employed were 3.9, 4.0, 3.3 and 6.2 eV for Mn, Fe, Co, and Ni, respectively, identical to the ones used in Materials Project.[35] The chemical bonding information by means of COHP, and Mulliken and Löwdin charge population analysis were performed using the LOBSTER package.[36, 37]

The general Li intercalation process into the bisulfate host structure (monoclinic or orthorhombic) can be represented by the following chemical reaction,

$$\mathrm{xLi^+ + xe^- + Li_yM(SO_4)_2 \rightarrow Li_{x+y}M(SO_4)_2}$$

where the $Li_{x+y}M(SO_4)_2$ and $Li_yM(SO_4)_2$ represent the lithiated and delithiated compositions, respectively. The average intercalation voltage is related to the difference in the Gibbs energies of the lithiated and delithiated phases, as given by the equation below.[38, 39]

$$V = -\frac{\left(E_{\mathrm{Li_{x+y}M(SO_4)_2}} - E_{\mathrm{Li_yM(SO_4)_2}} - \mathrm{x}E_{\mathrm{Li}}\right)}{\mathrm{x}F}$$

The intercalation voltage, as calculated by us, approximates the Gibbs energies with the total DFT energies (or $G \approx E$), which ignores the entropic and $p - V$ contributions. $E_{\mathrm{Li}}$ is the total energy of the body-centred-cubic structure of pure Li and $F$ is the Faraday constant. Note that we have considered only topotactic intercalation reactions in our study, i.e., we have used the calculated DFT energies of lithiated and delithiated structures which are initialized with the same monoclinic or orthorhombic symmetry.

While we obtained the fully intercalated monoclinic and orthorhombic structures from the inorganic crystal structure database (ICSD[40]) for the Fe, Co, and Ni bisulfates, we created the o-$Li_2Mn(SO_4)_2$ structure via ionic substitution of Co with Mn in o-$Li_2Co(SO_4)_2$. We formed the initial delithiated structures by removing Li atoms from the unrelaxed $Li_2M(SO_4)_2$ hosts, resulting in two distinct compositions, namely $LiM(SO_4)_2$ and $M(SO_4)_2$. In the case of $LiM(SO_4)_2$, several Li-vacancy configurations are possible, and we used the pymatgen package to enumerate all symmetrically distinct orderings.[41] We considered only the DFT-calculated lowest energy $LiM(SO_4)_2$ structures for subsequent calculations and analysis.

The charge density difference isosurfaces were calculated as follows: *i*) we removed one Li atom from a SCAN+*U*-relaxed structure and performed a single self-consistent-field (SCF) calculation, and *ii*) the charge density that was obtained from this SCF calculation was subtracted from the charge density of the original SCAN+*U*-relaxed structure. For example, we calculated the charge density difference isosurface between $Li_2Ni(SO_4)_2$ and $Li_{1.5}Ni(SO_4)_2$ by removing one Li per unit cell from the SCAN+*U*-relaxed $Li_2Ni(SO_4)_2$ and performing a single SCF calculation on the resultant Li-removed structure. We followed a similar process for calculating the charge density difference between $LiNi(SO_4)_2$-$Li_{0.5}Ni(SO_4)_2$ compositions.

## 4 Results

### 4.1 Phase stability and lattice parameters

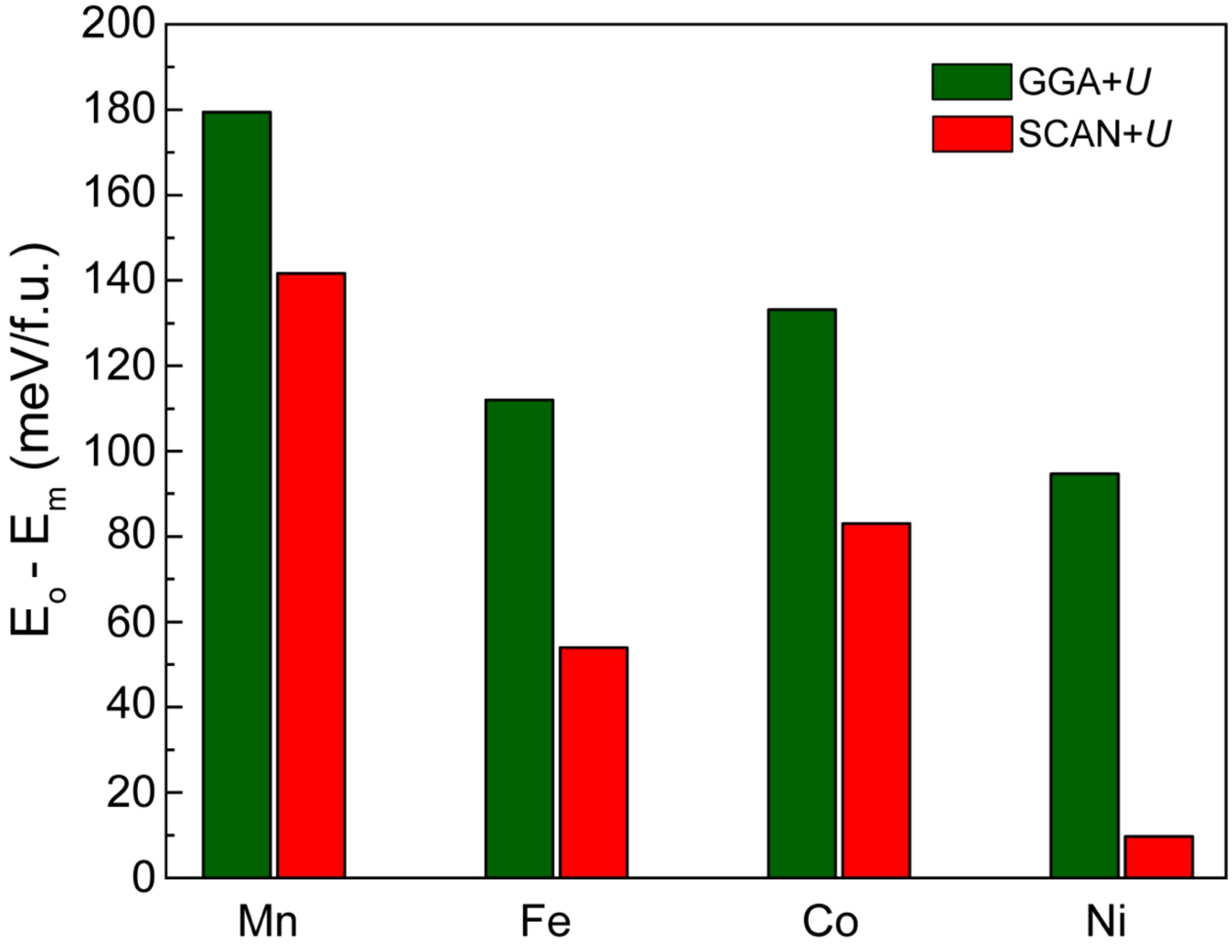

**Figure 2**: Difference in energy per formula unit between orthorhombic ($E_o$) and monoclinic ($E_m$) polymorphs of $Li_2M(SO_4)_2$ (M= Mn, Fe, Co, and Ni) using GGA+*U* (green bars) and SCAN+*U* (red bars) functionals.

**Figure 2** plots the energy difference (in meV per formula unit) between the monoclinic and orthorhombic polymorphs of $Li_2M(SO_4)_2$ (M = Mn, Fe, Co, and Ni), calculated using both GGA+*U* and SCAN+*U* functionals. The monoclinic phase is predicted to be more stable than the orthorhombic phase for all transition metals, with the Ni-analogue exhibiting the smallest energy difference between the two polymorphs (~9.7 meV/f.u. with SCAN+*U*). Previous phase stability measurements, via formation and dissolution enthalpies, using isothermal acid solution calorimetry analysis, are in agreement with our predictions for Co and Fe, while for Ni, the orthorhombic phase is found to be more stable experimentally.[18, 22] This disparity in the Ni-analogue may be due to the actual energy difference between the two polymorphs being within the experimental error bar or the neglect of entropy contributions within our theory framework. For the rest of the manuscript, we have considered the stable monoclinic phase for the cases of Mn, Fe, and Co, while we performed calculations in both polymorphs for the Ni analogue.

**Table 1**: GGA+*U*, SCAN+*U*, and experimental lattice parameters of m-$Li_2M(SO_4)_2$ (M = Mn, Fe, Co, and Ni). Vol. represents the lattice volume per conventional cell.

| System | | a (Å) | b (Å) | c (Å) | α (°) | β (°) | γ (°) | Vol. (Å$^3$) |
|---|---|---|---|---|---|---|---|---|
| $Li_2Mn(SO_4)_2$ | GGA+*U* | 5.0410 | 8.42716 | 8.99251 | 90.00 | 121.0298 | 90.00 | 327.808472 |
| | SCAN+*U* | 4.97516 | 8.26481 | 8.81151 | 89.9996 | 121.2164 | 90.00 | 309.860278 |
| | Expt. | 4.99198 | 8.33959 | 8.86140 | 90.00 | 121.2263 | 90.00 | 315.464260 |
| $Li_2Fe(SO_4)_2$ | GGA+*U* | 5.05012 | 8.27746 | 8.97427 | 89.9999 | 121.9008 | 90.00 | 318.483975 |
| | SCAN+*U* | 4.98604 | 8.11787 | 8.77091 | 90.0008 | 121.9297 | 90.0008 | 301.297131 |
| | Expt. | 4.98360 | 8.19100 | 8.81080 | 90.00 | 121.9150 | 90.00 | 305.293705 |
| $Li_2Co(SO_4)_2$ | GGA+*U* | 5.02342 | 8.18210 | 8.90553 | 90.00 | 121.5011 | 90.00 | 312.093725 |
| | SCAN+*U* | 4.94928 | 8.01501 | 8.72290 | 89.9999 | 121.3495 | 89.9999 | 295.507830 |
| | Expt. | 4.96710 | 8.09080 | 8.76390 | 90.00 | 121.8550 | 90.00 | 299.155587 |
| $Li_2Ni(SO_4)_2$ | GGA+*U* | 5.0098 | 8.10859 | 8.85295 | 89.9999 | 121.6909 | 90.001 | 305.99858 |
| | SCAN+*U* | 4.94941 | 7.92365 | 8.68075 | 89.9999 | 121.7791 | 89.9998 | 289.400177 |
| | Expt. | 4.96004 | 8.01217 | 8.72062 | 90.00 | 121.8880 | 90.00 | 294.260868 |

**Table 1** and **Table S1** of the supplementary information (SI) compile the lattice parameters and volumes of all the monoclinic and orthorhombic structures, respectively, as calculated by GGA+*U* and SCAN+*U*, as well as from experiments.[16, 17, 20, 22] **Tables S2** and **S3** tabulate the bond lengths, bond angle variances, and other information related to coordination polyhedra within the orthorhombic and monoclinic structures. Notably, both GGA+*U* and SCAN+*U* calculations are in good agreement with experimental values, with GGA+*U* (SCAN+*U*) overestimating (underestimating) lattice parameters by a maximum of 1% (0.1%), and are also consistent with previous theoretical studies.[42] Thus, we find that SCAN+*U* is quantitatively in marginally better agreement with experiments than GGA+*U*. In terms of lattice volumes, we observe that the volume of the conventional cell, for both monoclinic and orthorhombic phases, increase progressively with increasing atomic radius of the TM, a trend that is captured by both GGA+*U* and SCAN+*U*. Similar to the trend of lattice volumes, M-O bond lengths (see **Tables S2** and **S3**) also increase with increasing TM atomic radius, while S-O bond lengths remain roughly constant for all TM analogues.

### 4.2 Intercalation Voltages

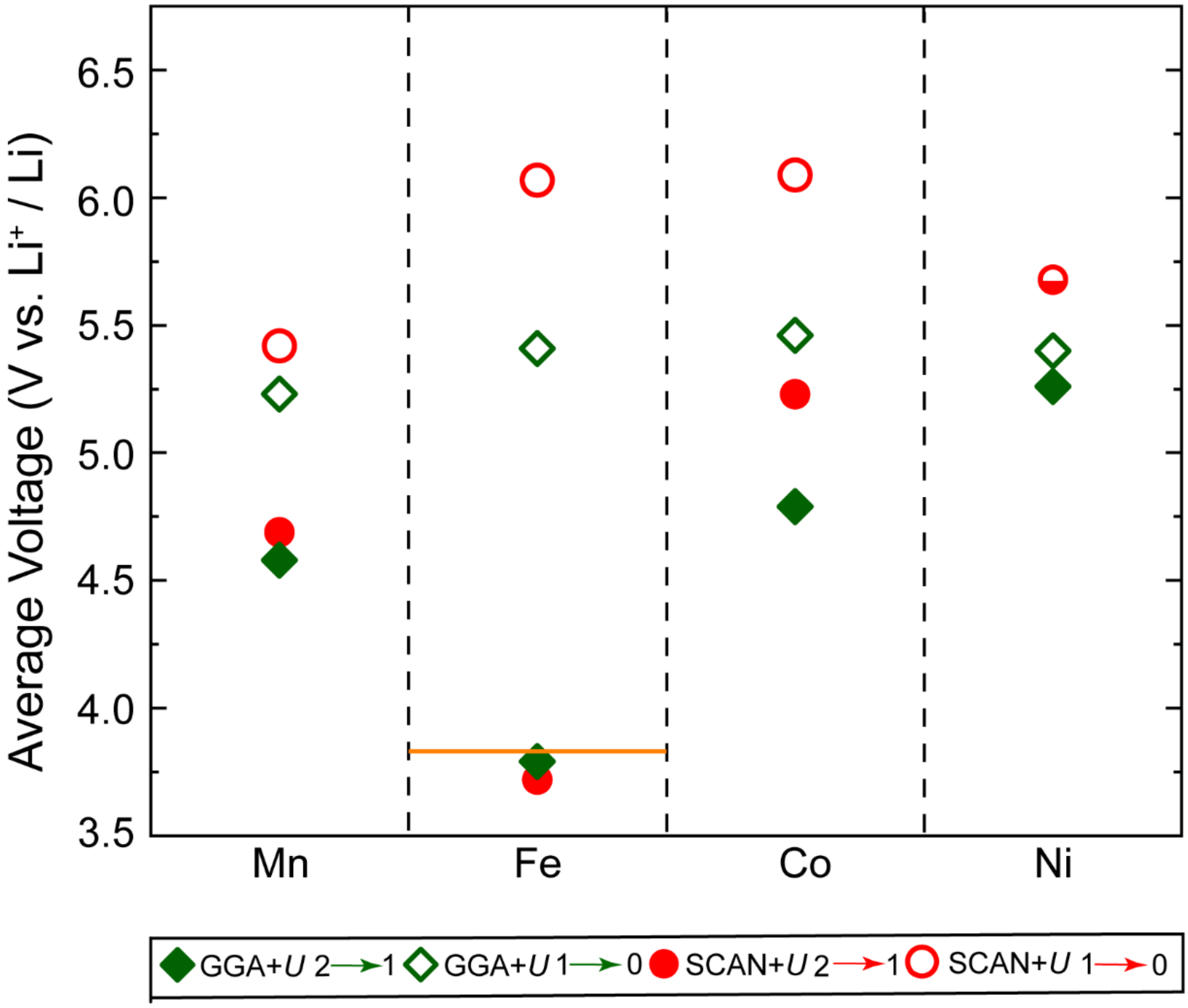

**Figure 3**: Average topotactic voltages of monoclinic m-$Li_xMn(SO_4)_2$ (M= Mn, Fe, Co, and Ni), as calculated by GGA+*U* (green diamonds) and SCAN+*U* (red circles) functionals. Filled and empty shapes represent the first ($1 \leq x \leq 2$) and second ($0 \leq x \leq 1$) deintercalation of Li from the host, respectively. Horizontal orange line represents experimental voltage. Half-filled red circle for the Ni analogue represents the average voltage across the entire Li deintercalation range ($0 \leq x \leq 2$), calculated by SCAN+*U*.

**Figure 3** presents the GGA+*U* (green diamonds) and SCAN+*U* (red circles) calculated average topotactic voltages (in units of V vs. $Li^+$/Li), for the monoclinic-$Li_xM(SO_4)_2$ hosts ($0 \leq x \leq 2$). Solid symbols in **Figure 3** represent the voltage for the first Li deintercalation, i.e., for the $Li_2M(SO_4)_2 \rightarrow LiM(SO_4)_2$ process or $1 \leq x \leq 2$, while hollow symbols represent the voltage for the second Li removal ($LiM(SO_4)_2 \rightarrow M(SO_4)_2$ or $0 \leq x \leq 1$). In the case of SCAN+*U* calculations in Ni bisulfate, we found the $LiNi(SO_4)_2$ monoclinic phase to be metastable relative to phase separation into $Li_2Ni(SO_4)_2$ and $Ni(SO_4)_2$, which explains the half-filled red circle for Ni in **Figure 3**. The experimental voltage for $Li_2Fe(SO_4)_2 \rightarrow LiFe(SO_4)_2$ is represented by the solid orange line in **Figure 3**.[16] Importantly, we observe that the SCAN+*U* calculated voltages are typically higher than the corresponding GGA+*U* values, consistent with recent trends reported in layered Li-TM-oxides.[43] Additionally, the average voltages across the entire Li concentration (i.e., $Li_2M(SO_4)_2 \rightarrow M(SO_4)_2$) is quite "high" for all TMs considered (range of average voltages: 4.7-5 V with GGA+*U* and 5-5.5 V with SCAN+*U*), which is consistent with trends reported for sodium sulfate cathodes as well.[44]

Our data with both GGA+*U* and SCAN+*U* functionals indicate that the voltage for first Li removal is higher for Mn, Co, and Ni analogues compared to Fe, which can be attributed largely to the stability of the 3+ oxidation state of Fe over its +2 state.[45] Mn, Co, and Ni bisulfates exhibiting higher intercalation voltages than the corresponding Fe-analogue is a known observation among polyanion cathodes as well.[46] While SCAN+*U* calculated voltages for Mn, Co, and Ni follows the trend of standard reduction potentials of the respective TMs, similar to observations in layered Li-TM-oxides,[43] GGA+*U* displays a qualitatively inconsistent trend of increasing voltages going from Mn to Ni. However, GGA+*U* is in better quantitative agreement (~3.79 V) with the known experimental voltage (~3.8 V) for $Li_2Fe(SO_4)_2 \rightarrow LiFe(SO_4)_2$ than SCAN+*U* (~4.3 V), also consistent with observations in Li-TM-oxides. Nevertheless, the overall lower self-interaction errors and lower magnitude of *U* corrections required with SCAN indicates that SCAN+*U* calculations will yield qualitatively better trends than GGA+*U*.

Although the voltage for first Li removal of Mn and Co bisulfates are higher than Fe, the voltage for second Li removal is the highest in the Fe compound (~6.6 V with SCAN+*U*) compared to all the other TMs (using SCAN+*U*), which can be attributed both to the instability of the $Fe^{4+}$ oxidation state and the increasing role played by anionic redox. Indeed, the second Li removal voltage is also significantly high for Co (~6.4 V, SCAN+*U*), which is symptomatic of anionic redox. The increase in voltage, in GGA+*U* calculations, going from first Li to second Li removal (~0.77 V on average across the 4 TMs) is lower than SCAN+*U* (~ 1.21 V on average across Mn, Fe, and Co). In any case, the predicted high voltages for second Li removal across all systems (>5.2 V with GGA+*U* and >5.5 V with SCAN+*U*) signify that it will be challenging to accomplish high extents of delithiation in the bisulfates considered in our work within the stable operating window of conventional Li-electrolytes.

### 4.3 Electronic structure

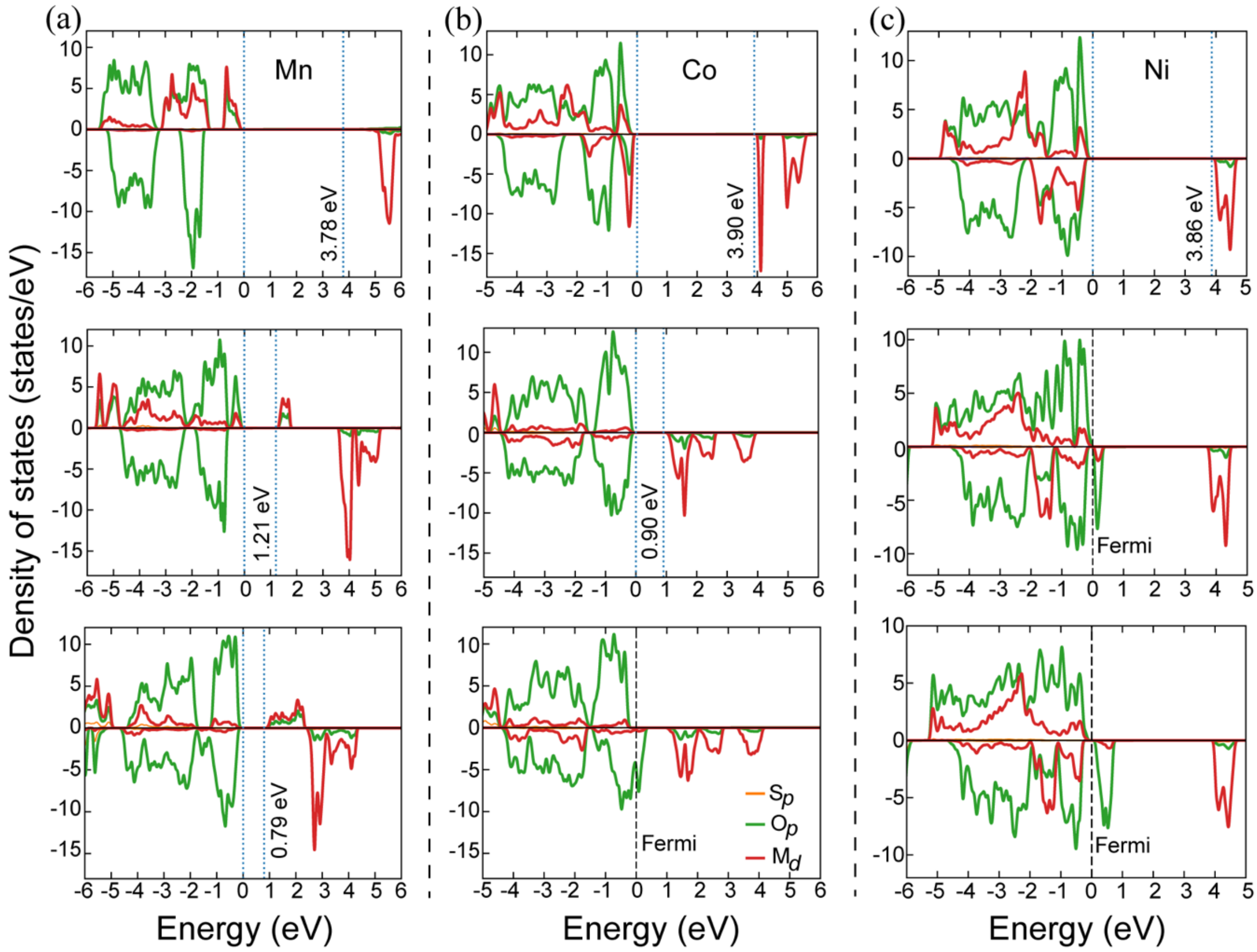

**Figure 4**: SCAN+$U$-calculated pDOS for m-$Li_2M(SO_4)_2$ (M = Mn, Co, and Ni in panels a, b, and c, respectively). The rows in each column indicate different states of lithiation, namely full lithiation ($x = 2$ in $Li_xM(SO_4)_2$) in the top row, half lithiation ($x = 1$) in the middle row, and no lithiation ($x = 0$) in the bottom row. Green, red and orange curves corresponds to O $p$, TM $d$, and S $p$ states respectively. Positive (negative) values of pDOS correspond to up (down) spin electrons. Dotted blue lines represent the valence and conduction band edges, with the numbers indicating band gap values. Dashed black lines signify Fermi level. The zero on the energy scale in each panel is referenced either to the valence band maximum or to the Fermi level.

The SCAN+$U$-calculated pDOS are plotted in **Figure 4** (and **Figure S2**) for the monoclinic structures of all TMs and in **Figure S3** and **S4** for the orthorhombic analogue. Panels a, b, and c in **Figure 4** represent Mn, Co, and Ni analogues, respectively, with orange, green, red, dotted blue, and dashed black lines representing S $p$ states, O $p$ states, M $d$ states, band edges, and calculated Fermi level, respectively. The top panels in **Figure 4** represented the fully lithiated ($Li_2M(SO_4)_2$) compositions, the middle panels highlight partially delithiated ($LiM(SO_4)_2$) states, and the bottom panels correspond to fully delithiated ($M(SO_4)_2$) structures. The general trend of "large" bandgaps predicted for all the lithiated compounds are consistent with observed values in polyanionic class of materials.[22] Although our GGA+$U$ band gaps (**Figure S5, S6,** and **S7**) are lower than SCAN+*U*, both functionals do predict qualitatively similar trends in band gaps and the electronic states at the band edges.

The general features of the electronic structure obtained through both the functionals in fully lithiated $Li_2M(SO_4)_2$ are similar for both the polymorphs. The TM *d* states dominate the valence band edge in the fully lithiated Co (and Fe) system(s), as shown in **Figure 4b** and **Figure S2,** while both TM *d* and O *p* states contribute in equal measure in $Li_2Mn(SO_4)_2$, indicating robust hybridization across the Mn-O bonds (**Figure 4a**). In $Li_2Ni(SO_4)_2$, O *p* states are the dominant contributors to the valence band edge, with Ni *d* states gravitating towards lower energies in the valence band (**Figure 4c**). Thus, our pDOS calculations in the $Li_2M(SO_4)_2$ structures indicate higher propensity for O redox in the Ni compound compared to the Mn, Fe, and Co analogues, upon first Li removal. Moreover, Ni-O bond lengths (see compilation in **Table S13**) do not indicate any Jahn-Teller-like distortion of the $NiO_6$ octahedra upon delithiation from $Li_2Ni(SO_4)_2$ to $LiNi(SO_4)_2$, indicating the lack of Ni oxidation from $Ni^{2+}$ to Jahn-Teller-active $Ni^{3+}$ with delithiation.

With partial delithiation, we observe a significant drop in band gap in all the bisulfates considered. For example, the SCAN+*U* band gap drops from 3.78 eV to 1.21 eV in Mn, 3.06 eV to 1.86 eV in Fe, 3.90 eV to 0.90 eV in Co, and 3.86 eV to metallic behavior in Ni. While the TM *d* states still contribute to the valence band edge in $LiMn(SO_4)_2$, the corresponding contributions are quite negligible in $LiCo(SO_4)_2$, indicating increasing likelihood of anionic redox in the Co-structure with further removal of Li. Importantly, the reduction of band gaps in Mn and Co structures with Li-removal also coincides with a shift in TM *d* states from the valence band edge (in $Li_2M(SO_4)_2$) to the new conduction band edge (in $LiM(SO_4)_2$), indicating a predominantly cationic redox. Note that the qualitative trends in the shift of TM *d* and O *p* states with delithiation in the Fe bisulfate is similar to that of the Co analogue.

In $LiNi(SO_4)_2$, the states around the Fermi level are predominantly occupied by O *p* states with minor contributions from the Ni *d* states. Importantly, on comparing the pDOS of $Li_2Ni(SO_4)_2$ and $LiNi(SO_4)_2$, we find that the states that formed the conduction band edge in $Li_2Ni(SO_4)_2$ (~4 eV above the valence band edge) remain at fairly similar energy levels with partial delithiation (~4 eV above the Fermi level), with the transition into metallic behavior largely arising from the shift of O *p* (and minor shift of Ni *d* states) from the valence band edge to energies higher than the Fermi level. Thus, our pDOS calculations provide a clear signature of anionic redox with (partial) delithiation in the Ni bisulfate structure. Note that the transition into metallic behavior also aides the delocalization of the hole across multiple O atoms, which can reduce the likelihood of peroxo-type bond formation and hence reduce $O_2$ gas evolution with Li removal. With O *p* states dominating the states that are slightly below the Fermi level, we can expect further contribution of anionic redox as more Li is removed to form the $Ni(SO_4)_2$ composition. Indeed, Ni-O bond lengths in $Ni(SO_4)_2$ (**Table S13**) do not exhibit any Jahn-Teller distortion which would indicate $Ni^{3+}$ formation.

Finally, with complete delithiation, the Co and Fe bisulfates transition into metallic behavior, while the Mn bisulfate remains semiconducting albeit with a smaller (0.79 eV) band gap compared to the partial delithiation structure. The transition into metallic behavior in Co and Fe structures exhibit signatures that are similar to the transition observed in $Li_2Ni(SO_4)_2 \rightarrow LiNi(SO_4)_2$. For example, in Co (and Fe), the states around the Fermi level are dominated by the O *p* states with marginal TM *d* contributions. Additionally, the metallic behavior arises largely due to the shift of O *p* states from the valence band edge of $LiCo(SO_4)_2$ (and $LiFe(SO_4)_2$) to energies higher than the Fermi level in $Co(SO_4)_2$ (and $Fe(SO_4)_2$). Thus, anionic redox dominates the second Li removal in Co and Fe bisulfates while TM redox dominates the first Li removal. The activation of anionic redox with second Li removal in Co and Fe is also consistent with the significantly higher voltages (> 6 V with SCAN+*U*, **Figure 3**) that we observe for second Li removal compared to the first Li removal (< 5.3 V with SCAN+*U*).

During the final Li removal, the signature of TM redox (than anionic redox) is strongest in the Mn bisulfate. For example, Mn *d* states dominate the conduction band edge (with minor contributions from O *p* states) in $LiMn(SO_4)_2$ as well as $Mn(SO_4)_2$, indicating that the holes with Li extraction are introduced predominantly in the TM *d* orbitals. Moreover, there is a significant peak in O *p* states with negligible Mn *d* states at the valence band edge in $Mn(SO_4)_2$, which has Mn in the +4 oxidation state (so any further oxidation will involve anionic redox), while Mn *d* states do contribute significantly in both $LiMn(SO_4)_2$ (where Mn is in +3 oxidation state and can reach the +4 state upon oxidation), and $Li_2Mn(SO_4)_2$ (Mn in +2 and can attain the +3 state upon oxidation). Also, Mn-O bond lengths (**Table S13**) exhibit clear Jahn-Teller distortion in $LiMn(SO_4)_2$ compared to Mn-O bonds in $Li_2Mn(SO_4)_2$ and $Mn(SO_4)_2$, signifying the formation of Jahn-Teller-active $Mn^{3+}$ upon first delithiation and non-Jahn-Teller-active $Mn^{4+}$ upon second delithiation. Thus, for the range of delithiation considered in our work, we don't expect any anionic redox to be active in the Mn bisulfate. In contrast, the anionic redox signature continues to remain strong with the final Li removal in the Ni analogue, with a larger population of O *p* states above the Fermi level, which signifies more holes introduced in O *p* orbitals than Ni *d*. Thus, based on our pDOS data, we can state that anionic redox is inactive in the Mn bisulfate, becomes progressively active in the Fe and Co analogues with increasing Li removal, and precedes any TM redox in the Ni structure.

### 4.4 On-site magnetic moments

To further probe the anionic redox activity in the Fe, Co, and Ni bisulfates, we analysed the SCAN+*U*-calculated on-site magnetic moments of the TM and O in each system, which are listed in **Tables S4-S8**. In the case of the Ni bisulfate, we observe that the change in the average magnetic moment of Ni, with SCAN+*U*, for the first and second Li removal are quite low (4% and -1.42% respectively), compared to the average change in magnetic moment of O atoms (505.36% and 0.90%, respectively), highlighting anionic redox. On the other hand, we find that only the average magnetic moment of Mn atoms changes significantly upon Li removal (-16.50% and -22.01% change with first and second Li extraction) in the Mn bisulfate, a signature of cationic redox across the range of delithiation considered. For the Co and Fe systems, changes in on-site magnetic moments are consistent with cationic redox for first Li removal followed by anionic redox for second Li removal, in agreement with our pDOS data (**Figure 4** and **Figure S2**).

### 4.5 Charge density difference and charge population

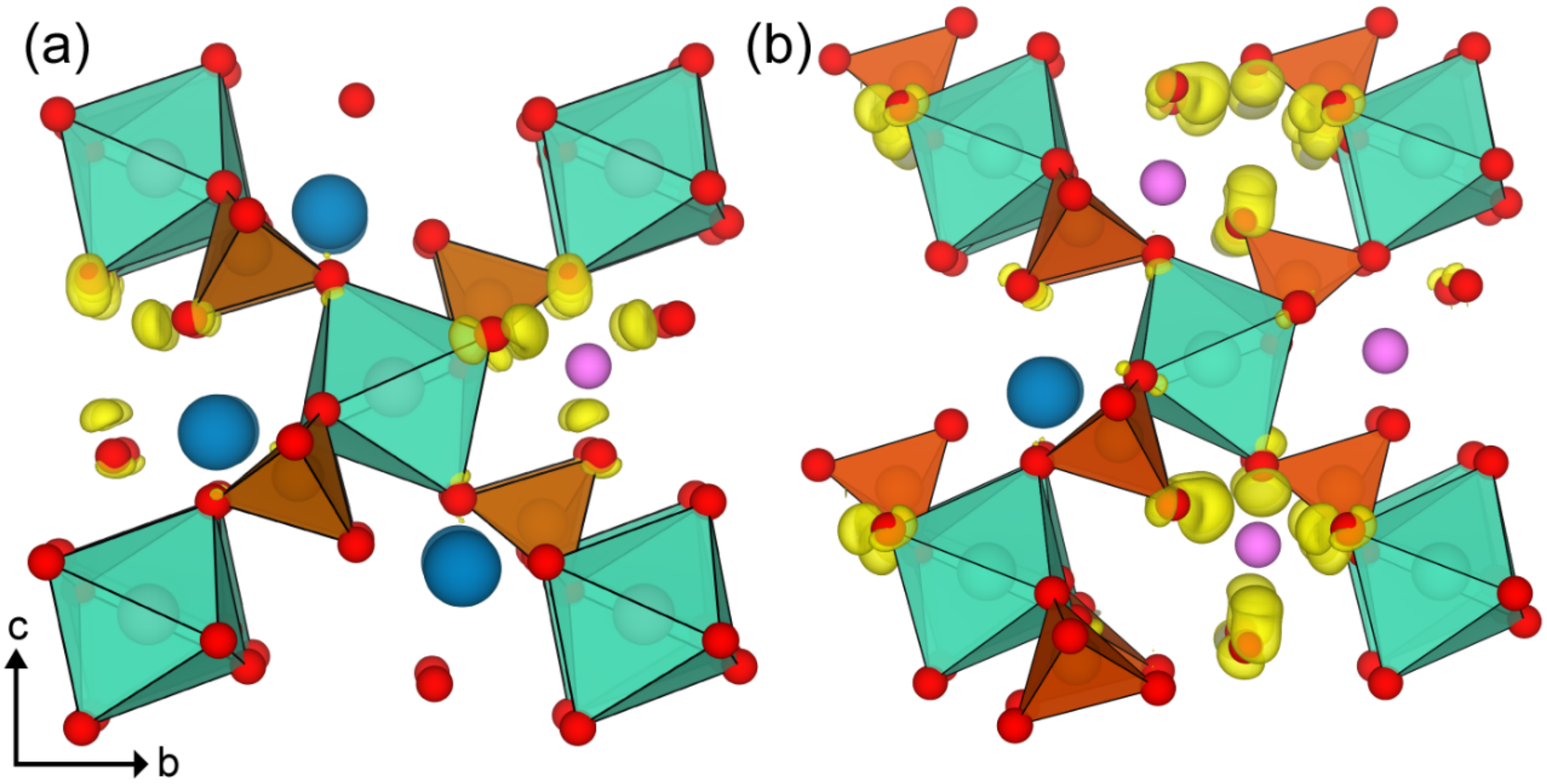

**Figure 5**: Differential charge density map for removal of one Li atom from relaxed unit cell of (a) $Li_2Ni(SO_4)_2$ (b) $LiNi(SO_4)_2$. The isosurfaces displayed (yellow lobes) correspond to a deficiency of 0.01 e/bohr$^3$. Pink sphere displays the removed Li atom, while blue spheres represent Li atoms that are present. Cyan and orange polyhedra represent $NiO_6$ and $SO_4$ groups, respectively, with oxygens represented by red spheres.

To visualize the physical regions where holes are (de)localized upon Li removal in bisulfates, we present the electronic charge density difference isosurfaces between $Li_2Ni(SO_4)_2$-$Li_{1.5}Ni(SO_4)_2$ and $LiNi(SO_4)_2$-$Li_{0.5}Ni(SO_4)_2$ pairs in panels a and b, respectively, of **Figure 5**. Notably, we find that the holes (yellow lobes in **Figure 5**) resulting from both first and second Li extraction (extracted Li atoms indicated by pink spheres) exclusively occur in the vicinity of nearby O atoms, with a minor degree of delocalization on Ni atoms, which is definitive evidence for anionic redox preceding any Ni redox. For the Co bisulfate, the charge density difference isosurfaces (**Figure S8**) suggest a largely cationic redox upon first Li removal, with holes localized largely on Co atoms. Subsequent Li extraction in $LiCo(SO_4)_2$ results in a charge density difference isosurface that appears similar to the Ni analogue, signifying anionic redox. The evidence for anionic redox in the Ni and Co bisulfates is further corroborated by the Mulliken and Löwdin charge population analysis, as compiled in **Tables S9-S12**. The unusual hole delocalization that we observe in the case of Mn bisulfates (**Figure S8**) can be attributed to the Jahn-Teller distortion associated with the $Mn^{3+}$ state (see **Discussion** section for further description).

## 4.6 COHP

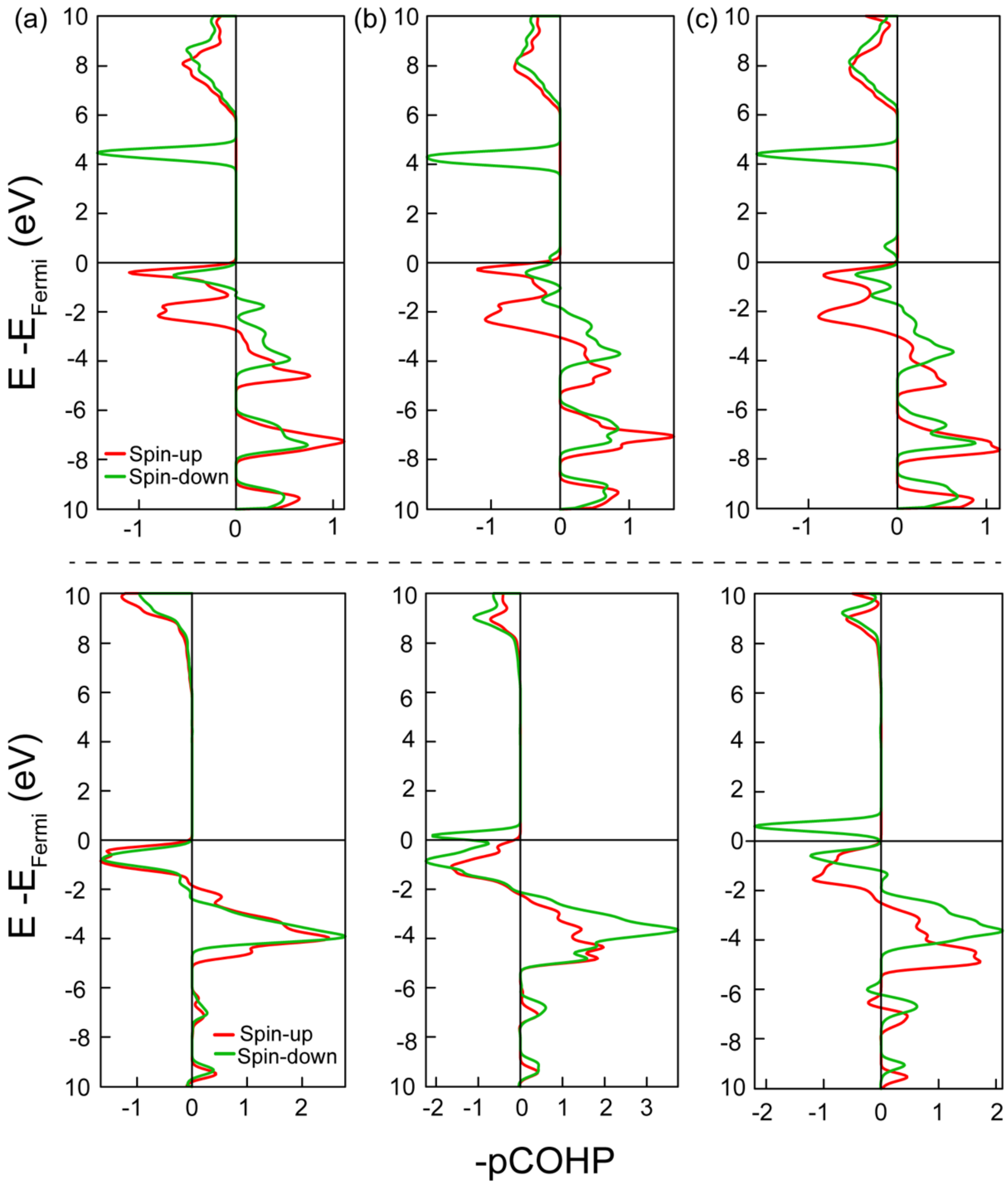

**Figure 6**: Negative projected crystal orbital Hamiltonian population of Ni-O (top row) and O-O (bottom row) bonds in (a) fully lithiated $Li_2Ni(SO_4)_2$, (b) half-delithiated $LiNi(SO_4)_2$, and (c) fully delithiated $Ni(SO_4)_2$ monoclinic structures. Positive and negative values on the horizontal axis indicate bonding and antibonding interactions, respectively.

To examine the possibility of irreversible $O_2$ evolution with anionic redox in Ni bisulfate, we inspect possible bonding of $O_2^{n-}$ species by monitoring the changes in the negative projected COHP (-pCOHP) of O and Ni atoms that neighbour the O atoms that significantly change their on-site magnetic moments with Li removal in monoclinic structures of $Li_xNi(SO_4)_2$, as displayed in **Figure 6**. Top (bottom) row of **Figure 6** represents the Ni-O (O-O) bonds, while each column represents different extent of lithiation in the Ni bisulfate. Notably, in both $Li_2Ni(SO_4)_2$ and $LiNi(SO_4)_2$ (panels a and b in **Figure 6**), antibonding states dominate near the Fermi level, revealing that the antibonding electrons are withdrawn during the oxidation. Importantly, there is no substantial increase in either bonding or antibonding states near the Fermi level in both Ni-O as well as O-O bonds across the extent of delithiation, indicating weak interactions among neighboring O atoms and suggesting low chances of $O_2$ evolution. Thus, we don't expect $O_2$ release to be a significant challenge with anionic redox in the bisulfates considered in our work, which can be attributed to the highly stable, covalent S-O bonds that form the structural framework.

## 5 Discussion

In this work, we have used Hubbard *U* corrected DFT calculations to explore the polymorph stability, Li intercalation voltage, and the participation of cationic and/or anionic redox in Mn, Fe, Co, and Ni containing monoclinic (and orthorhombic) bisulfate frameworks. Importantly, our calculated pDOS, on-site magnetic moments, charge density difference isosurfaces, Mulliken and Löwdin charge populations, and COHP point to anionic redox dominating throughout delithiation in $Li_2Ni(SO_4)_2$. In the case of Co and Fe bisulfates, we observe cationic redox to be predominant during the first Li removal ($Li_2M(SO_4)_2 \rightarrow LiM(SO_4)_2$), while anionic redox dominates the second Li removal. We did not find any evidence for anionic redox in the Mn analogue. Finally, our COHP data also points to low likelihood for $O_2$ release during anionic redox in the bisulfates, which is promising for the eventual reversibility of the anionic redox.

A common observation in the bisulfates considered in this work is that the computed average intercalation voltages (**Figure 3**) are beyond the anodic stability of conventional Li liquid electrolytes, with the exception of the $Li_2Fe(SO_4)_2 \leftrightarrow LiFe(SO_4)_2$ redox. Thus, experimental techniques that can probe the electronic structure can be used in place of electrochemical measurements to examine the properties of $Li_2M(SO_4)_2$ and validate our theoretical predictions. For example, techniques like X-ray absorption spectroscopy or X-ray photon spectroscopy can be used to probe the electronic states of the species present in the bisulfate framework, which can then be compared with the oxidation states predicted in our work. Further, optical band gap measurements can be used to qualitatively check the accuracy of our predictions as well.

However, we note that the application of the above mentioned experimental techniques to study bisulfates may have practical constraints as well.

Similar to trends observed in Li-TM-oxide cathodes, we observe our GGA+*U*-calculated voltage in the Fe bisulfate to be quantitatively more accurate with the experimental value than the SCAN+*U*-calculated voltage.[43] In general, we observe SCAN+*U*-calculated voltages to be higher than the corresponding GGA+*U*-calculations, also in agreement with Li-TM-oxide trends.[43] The overestimation of intercalation voltages with SCAN+*U* can be attributed to the known underestimation of energies (i.e., total energies are less negative) of metastable and unstable phases by SCAN+*U* compared to GGA+*U*.[43] Also, our SCAN+*U*-predicted band gaps are higher than GGA+*U*, and given DFT is a ground-state theory, we expect our SCAN+*U* band gaps to be in better agreement with any future experimental measurements. In terms of polymorph stability, lattice parameters, and qualitative trends of band gaps, we find both SCAN+*U* and GGA+*U* functionals to be in agreement with each other (and with available experimental data), which augurs well for the rest of our theoretical predictions. While hybrid functionals can be used to probe anionic redox, as has been done in other Ni-based oxides,[4] we did not perform any hybrid functional calculations given its high computational cost. Nevertheless, we do not expect any qualitative disagreement to arise even if calculations are performed with hybrid functionals in the future.

In the case of Mn bisulfates, we observe an unusual extent of hole delocalization (**Figure S8**), which is in contrast to the observed band gaps in pDOS (**Figure 4a**) and the on-site magnetic moments (**Tables S4** and **S6**). This hole delocalization behaviour in Mn bisulfates is not physically precise and can be attributed to the Jahn-Teller distortion of $Mn^{3+}$ and the fact that we do not allow for any ionic relaxation when we calculate charge density differences between two structures. Hence, the lack of Jahn-Teller distortion, in the case of charge density difference of $Li_2Mn(SO_4)_2 \rightarrow LiMn(SO_4)_2$, and the excess Jahn-Teller distortion (in $LiMn(SO_4)_2 \rightarrow Mn(SO_4)_2$) causes the unphysical hole delocalization.

Previous work has reported that an electron transfer from O *p* to M *d* states, through a so-called reductive coupling mechanism,[5] can trigger irreversible superoxide formation and eventual release of $O_2$ molecules from the lattice whenever anionic redox is active. Specifically, the charge transfer from O *p* to M *d* states increases the hole concentration on oxygen atoms, which facilitates the formation of peroxo bonds ($O_2^{n-}$) and eventually $O_2$ molecule. In the case of $Ni(SO_4)_2$, we observe only minor reductive coupling across the Ni-O bonds, signified by on-site magnetic moment changes on Ni atoms of ~0.026 on average (**Table S4**). Note that, this magnitude of electron transfer to Ni *d* states is insufficient to promote $O_2^{n-}$ bonding as it requires the transfer of ~1e$^-$ per Ni. Thus, the lack of reductive coupling further validates our hypothesis that anionic redox in bisulfate frameworks will not lead to peroxo formation and $O_2$ evolution, consistent with our -pCOHP data (**Figure 6**).

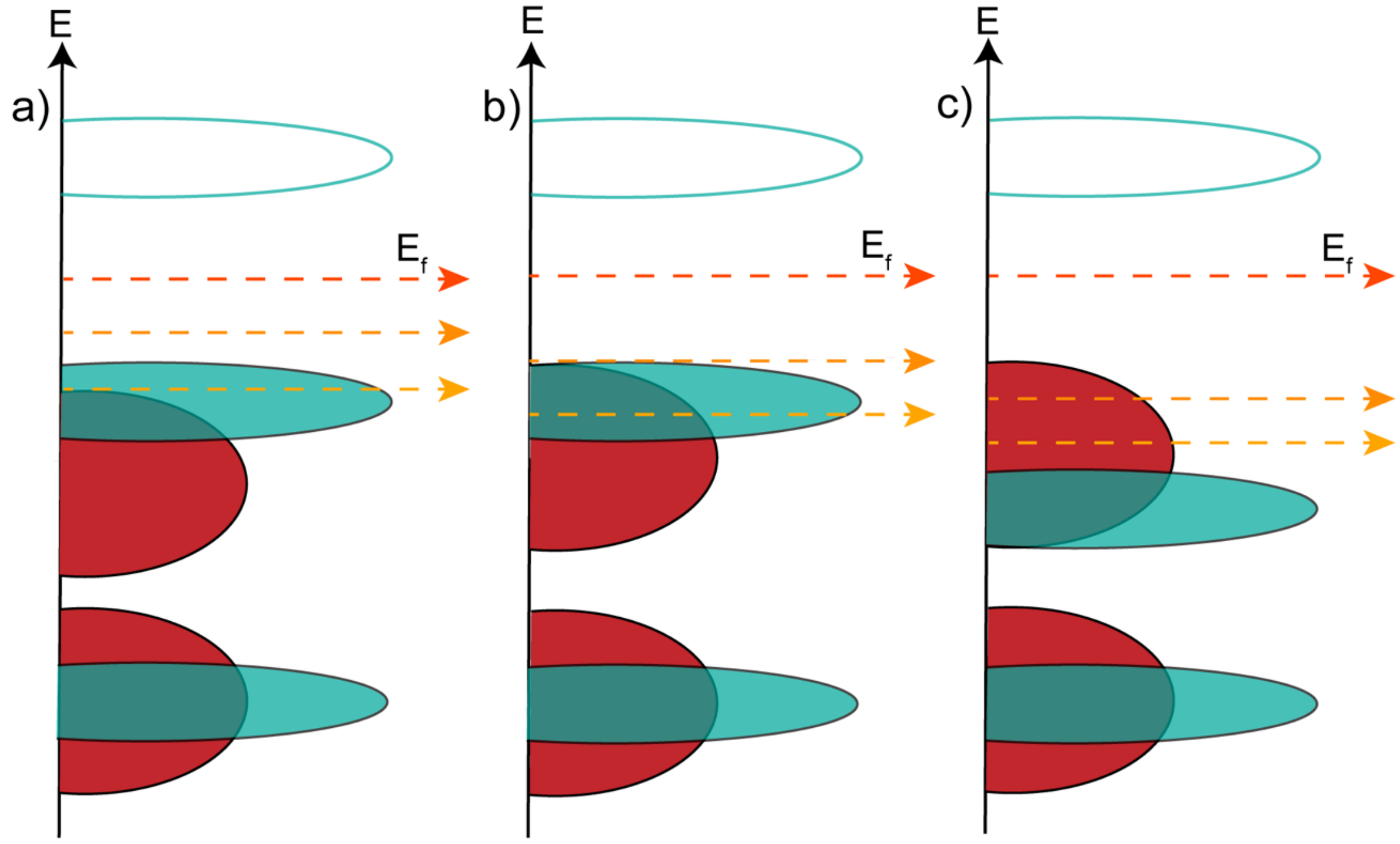

**Figure 7**: Illustration of the different distributions of transition metal *d* (blue lobes) and O *p* (red lobes) states in the bisulfates considered in our work, where in transition metal states are higher, similar, and lower in energy than oxygen states, in panels a, b, and c, respectively. The dashed red, orange, and yellow lines indicate progressive reduction in the Fermi level ($E_f$) with progressive delithiation in the bisulfates.

Finally, we provide a unified DOS schematic of the redox mechanism in **Figure 7** to better illustrate the participation of cationic and anionic redox processes in the bisulfates considered in this work. Red, filled blue, and hollow blue indicate filled O *p* states, filled M *d* states, and empty M *d* states, as would be the typical case in the bisulfate frameworks. Thus, upon delithiation, electrons from states that are closest to the Fermi energy ($E_f$ in **Figure 7**) will be removed, which indicates whether the electrons are removed from M (cationic redox) or O (anionic redox). Note that $E_f$ will drop in energy as electrons are removed from the bisulfate framework, as indicated by the dashed red, orange, and yellow lines, which correspond to $E_f$ in $Li_2M(SO_4)_2$, $LiM(SO_4)_2$, and $M(SO_4)_2$ compositions, respectively.

For the Mn bisulfate, the Mn *d* states are closer to $E_f$ relative to the O *p* states, signifying pure cationic redox across the entire delithiation range, as qualitatively shown in **Figure 7a**. In the case of Co and Fe (**Figure 7b**), during first delithiation, the M *d* states are dominant below $E_f$ with marginal contributions from O *p* states as well, highlighting cationic redox. However, with the drop in $E_f$ upon Li removal, O *p* states increasingly contribute to the oxidation and eventually anionic redox dominates over cationic redox during the second delithiation. Finally, in the Ni bisulfate, the O *p* states are dominant below $E_f$ throughout the delithiation process, highlighting that anionic redox occurs preferably over cationic redox throughout.

## 6 Conclusions

Herein, we explored the anionic and/or cationic redox chemistry in transition metal bisulfates, which are important class of compounds for developing high voltage (and high energy density) Li-ion batteries, using both GGA+*U* and SCAN+*U* calculations. Specifically, we computed the polymorphic stability (between orthorhombic and monoclinic structures), average topotactic Li intercalation voltages, band gaps via pDOS, on-site magnetic moments, charge density difference isosurfaces, Mulliken and Löwdin charge populations, and the pCOHP in $Li_xM(SO_4)_2$ frameworks, where $0 \leq x \leq 2$ and M = Mn, Fe, Co, and Ni. Importantly, we found anionic redox to be active across all Li extraction in the Ni bisulfate, while cationic redox was active throughout in the Mn bisulfate. In the case of Fe and Co bisulfates, cationic redox dominated the first Li removal ($1 \leq x \leq 2$), while anionic redox dominated the second Li removal ($0 \leq x \leq 1$). The signature of anionic (or cationic) redox was verified using the SCAN+*U* calculated pDOS, M-O bond lengths, and on-site magnetic moments for all TM bisulfates. In the case of the Ni bisulfate, we also visualized the hole (de)localization with Li removal by tracking the charge density difference isosurfaces, which provided unambiguous evidence of anionic redox. Further, our pCOHP data indicated low susceptibility of peroxo and $O_2$ molecule formation in the Ni bisulfate, given the marginal amount of antibonding state overlap across neighbouring O atoms. While our GGA+*U*-calculated voltage provided better quantitative agreement with the experimental value in $Li_xFe(SO_4)_2$, we expect our SCAN+*U*-calculated band gaps to be in better qualitative agreement with experiments. Additionally, GGA+*U* and SCAN+*U* were in qualitative agreement in their lattice parameter and polymorph stability predictions, validating the rest of our theoretical predictions. We hope that our work provides a new dimension, especially in using the SCAN+*U* functional, to discover and understand novel high voltage intercalation cathodes for future generation of high energy density batteries.

**Acknowledgments**

G.S.G. acknowledges financially support from the Indian Institute of Science (IISc) Seed Grant, SG/MHRD/20/0020 and SR/MHRD/20/0013. P.K.J and S.S acknowledge the Ministry of Human Resource Development (MHRD, Government of India) for financial assistance. P.K.J would like to thank Prof. Abhishek Singh, Materials Research Centre, IISc for teaching the fundamentals of computational material science. P.B. thanks the Technology Mission Division (Department of Science and Technology, Government of India) for financial support under the Materials for Energy Storage program (DST/TMD/MES/2k18/207). We acknowledge the computational resources provided by the Supercomputer Education and Research Centre, IISc, for enabling some of the density functional theory calculations showcased in this work.

# Supporting Information

# First principles investigation of anionic redox in bisulfate lithium battery cathodes

*Pawan Kumar Jha,† Shashwat Singh,† Mayank Srivastava,$ Prabeer Barpanda† and Gopalakrishnan Sai Gautam*‡*

† Faraday Materials Laboratory (FaMaL), Materials Research Centre, Indian Institute of Science, Bangalore 560012, India

$ Department of Electronic Systems Engineering, Indian Institute of Science, Bangalore 560012, India

‡ Department of Materials Engineering, Indian Institute of Science, Bangalore 560012, India

**Corresponding Author**

*** E-mail:** saigautamg@iisc.ac.in

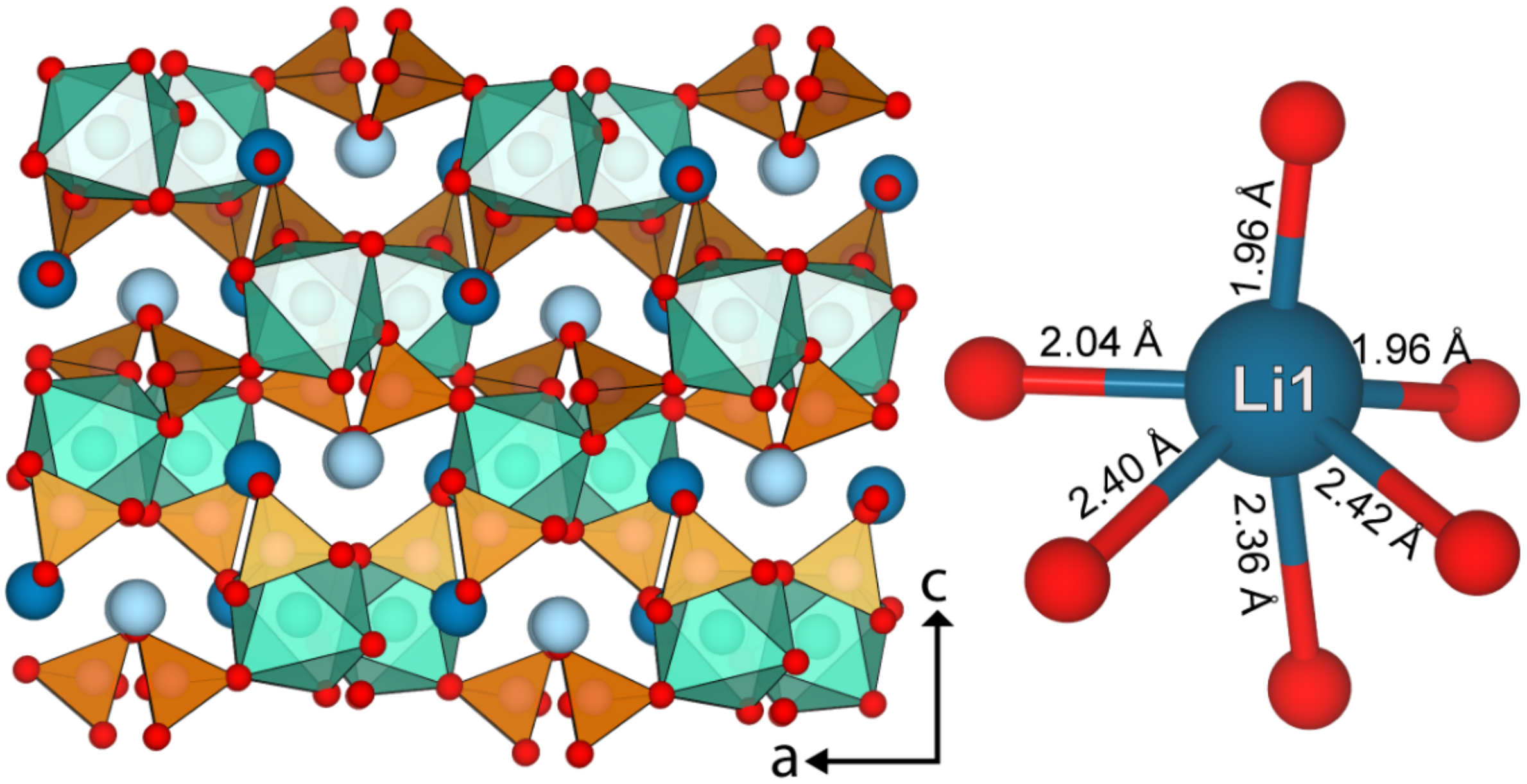

**Figure S1.** Crystal framework of *o*-$Li_2M(SO_4)_2$ along *b* direction. Dark and light blue spheres represent Li1 and Li2 sites. Orange and cyan polyhedra indicate the $SO_4$ tetrahedra and $MO_6$ octahedra. Right panel represents the distorton in $LiO_6$ octahedra of Li1 sites.

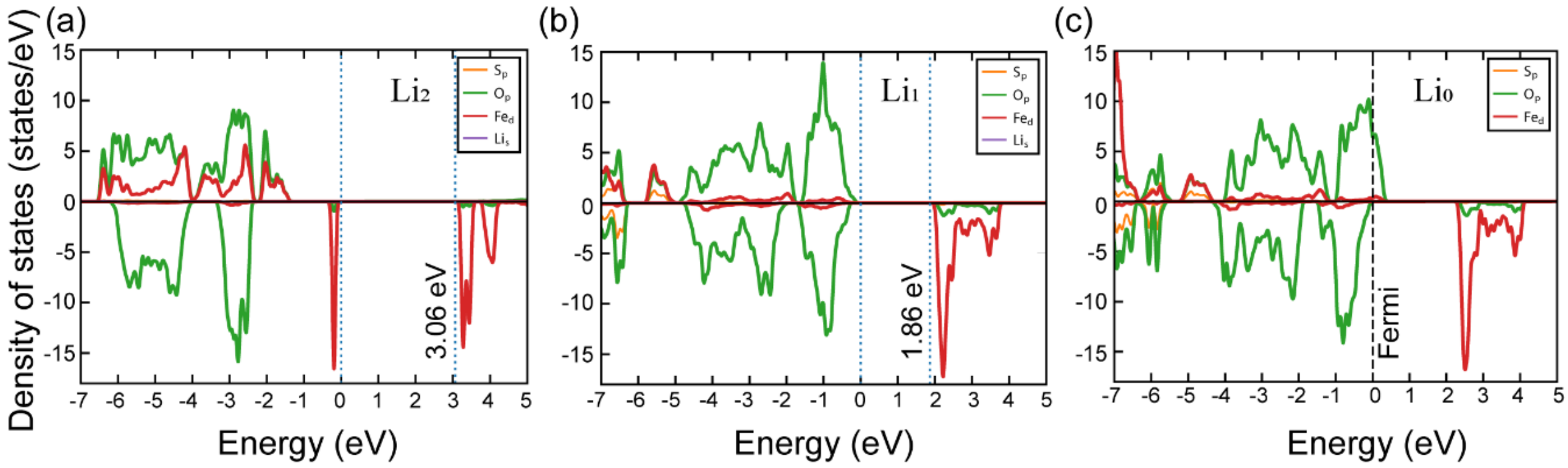

**Figure S2.** Projected density of state (pDOS) for monoclinic (*m*)-$Li_xFe(SO_4)_2$ at different states of (de)lithiation, namely x = 2, 1, and 0, as calculated by the Hubbard *U* corrected strongly constrained and appropriately normed (SCAN+*U*) functional. Green, red and orange curves correspond to O *p*, Fe *d*, and S *p* states respectively. Dotted blue lines represent the valence and conduction band edges, while dashed black line indicates the Fermi level. Band gaps are indicated by the text notations at the corresponding conduction band minima.

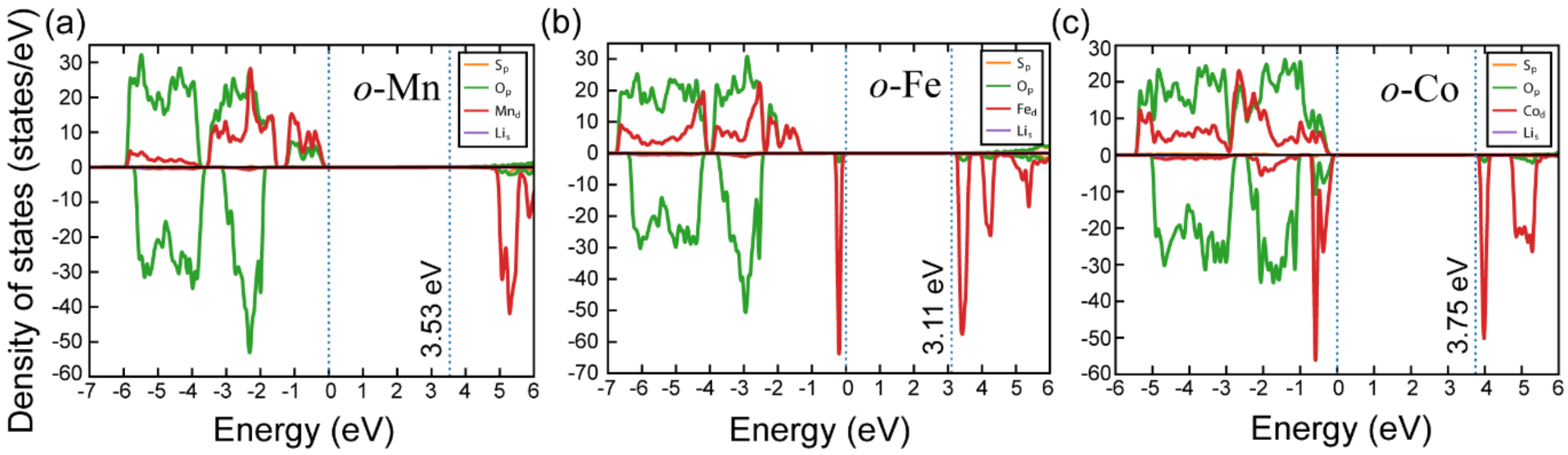

**Figure S3.** SCAN+*U*-calculated pDOS for orthorhombic (*o*)-$Li_2M(SO_4)_2$, where M = Mn, Fe and Co. Notations on each panel are identical to **Figure S2**.

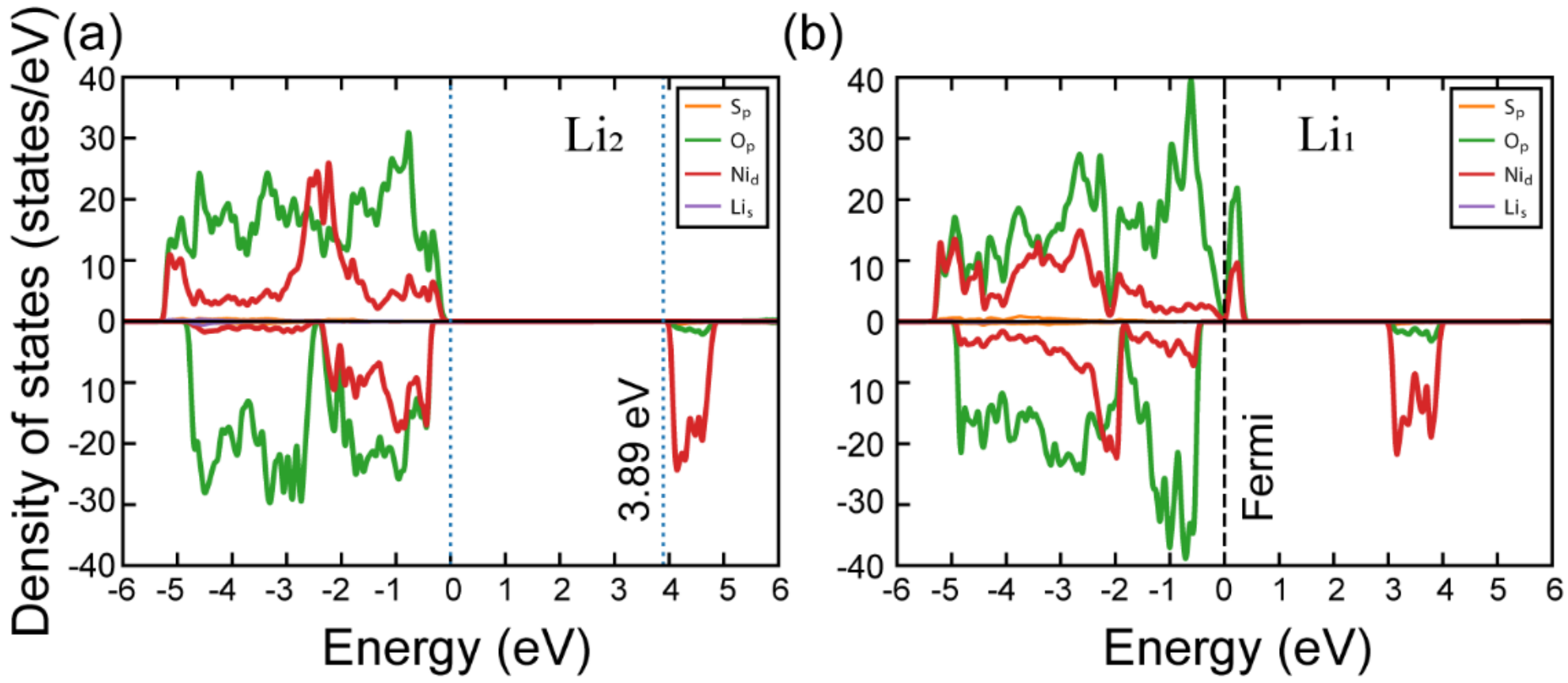

**Figure S4.** SCAN+*U*-calculated pDOS of *o*-$Li_xNi(SO_4)_2$ for x = 2 and 1. Notations on each panel are identical **Figure S2**.

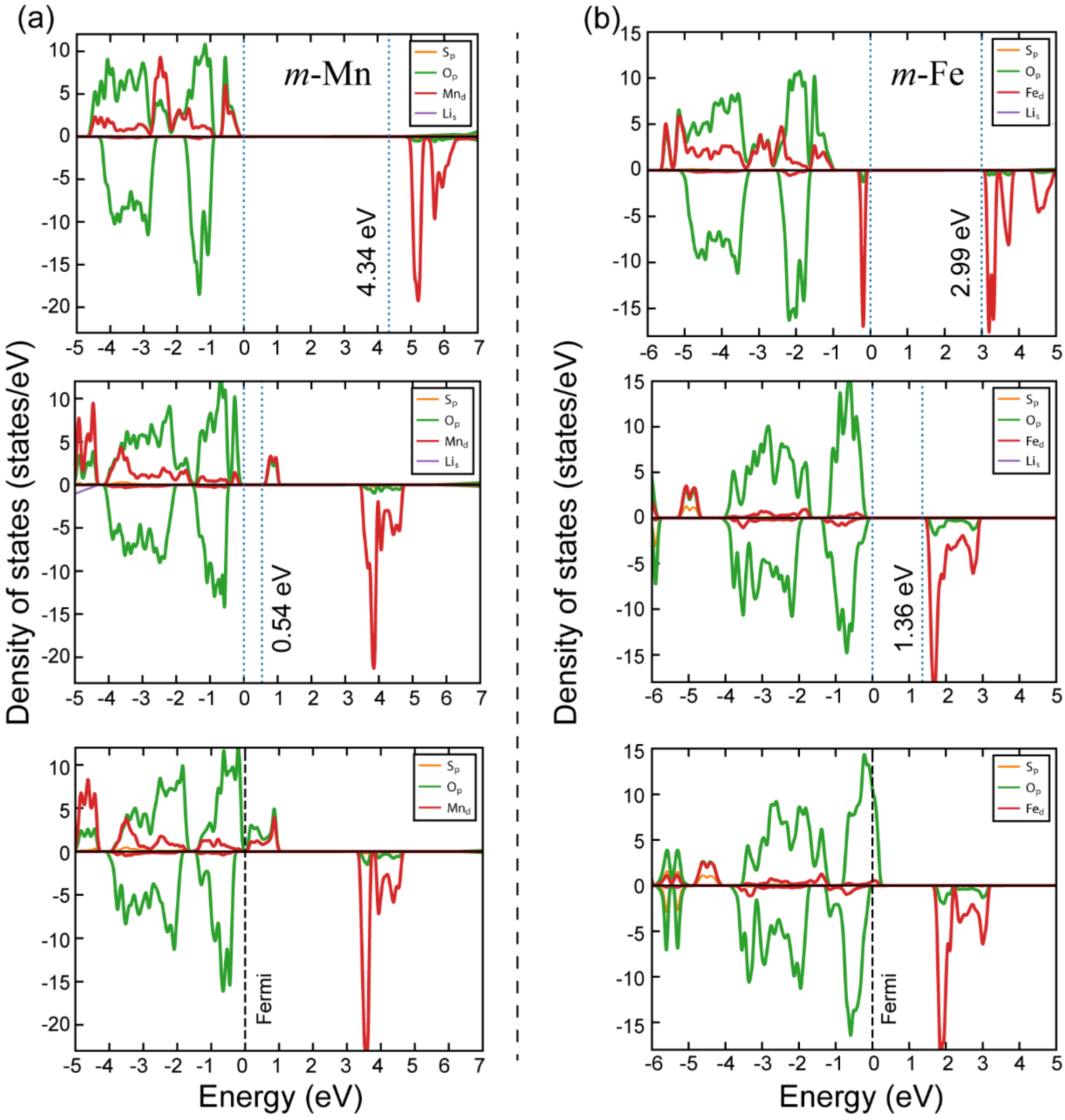

**Figure S5.** Hubbard *U* corrected generalized gradient approximation (GGA+*U*)-calculated pDOS of (a) *m*-$Li_xMn(SO_4)_2$ and (b) *m*-$Li_xFe(SO_4)_2$. Top, middle, and bottom panels represent different levels of (de)lithiation, namely, x = 2, 1, and 0, respectively. Notations on each panel are identical to **Figure S2.**

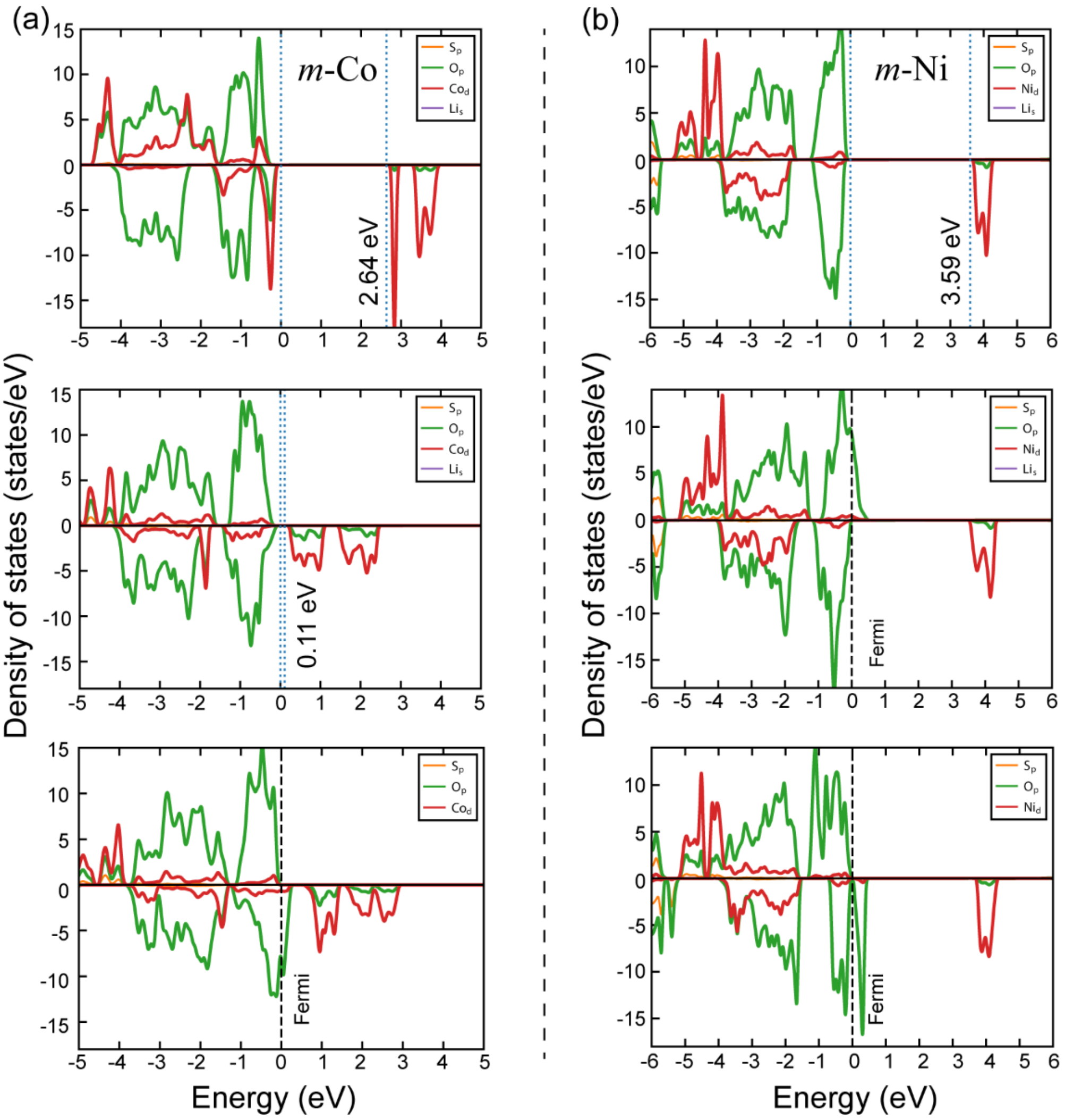

**Figure S6.** GGA+$U$-calculated pDOS of (a) $m$-$Li_xCo(SO_4)_2$ and (b) $m$-$Li_xNi(SO_4)_2$. Top, middle, and bottom panels represent different levels of (de)lithiation, namely, x = 2, 1, and 0, respectively. Notations on each panel are identical to **Figure S2.**

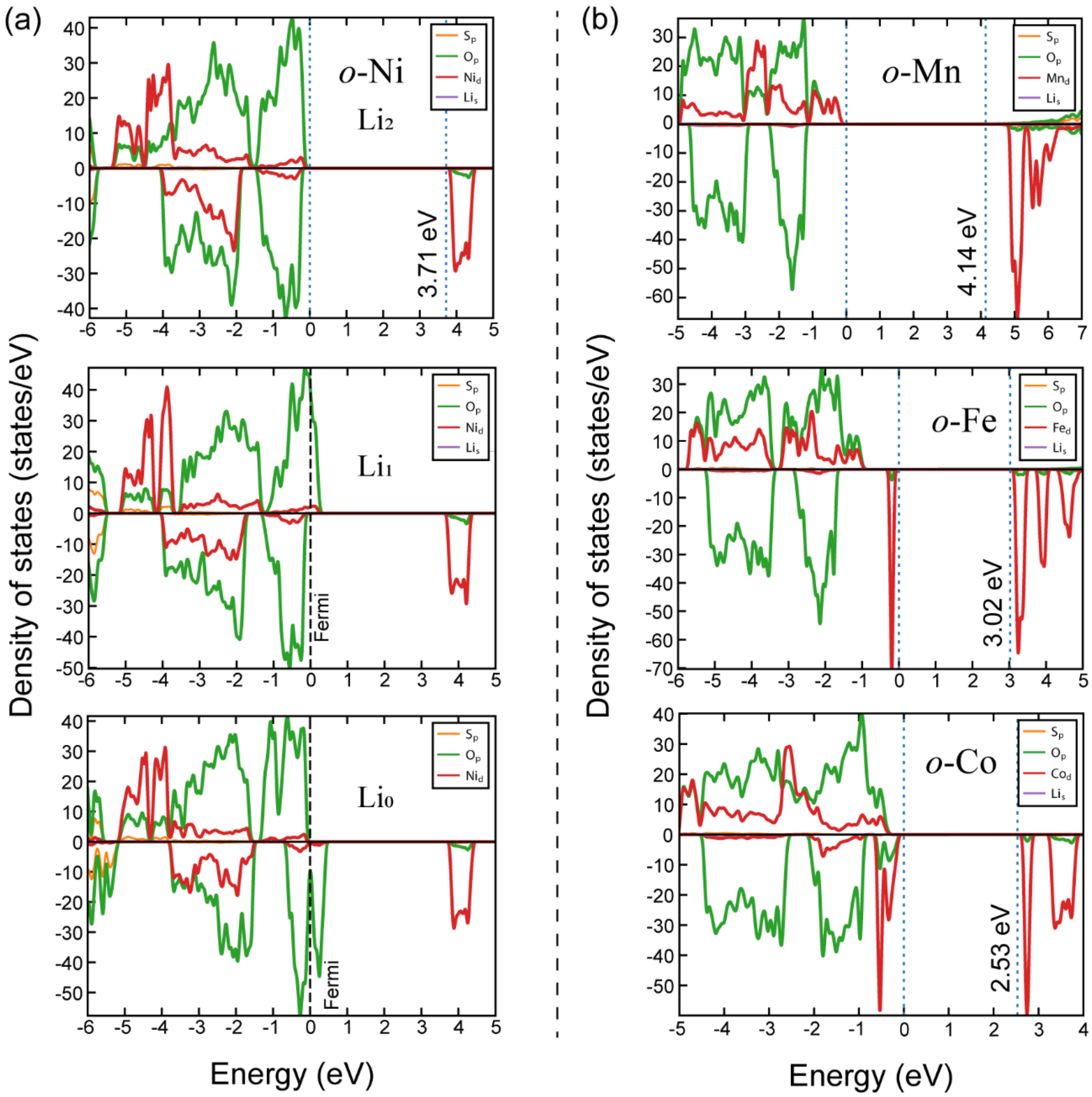

**Figure S7.** GGA+$U$-calculated pDOS of (a) $o$-$Li_xNi(SO_4)_2$ and (b) $o$-$Li_2M(SO_4)_2$, where M = Mn (top panel), Fe (middle), and Co (bottom). In (a), the top, middle, and bottom panels represent different levels of (de)lithiation, namely, x = 2, 1, and 0, respectively. Notations on each panel are identical to **Figure S2.**

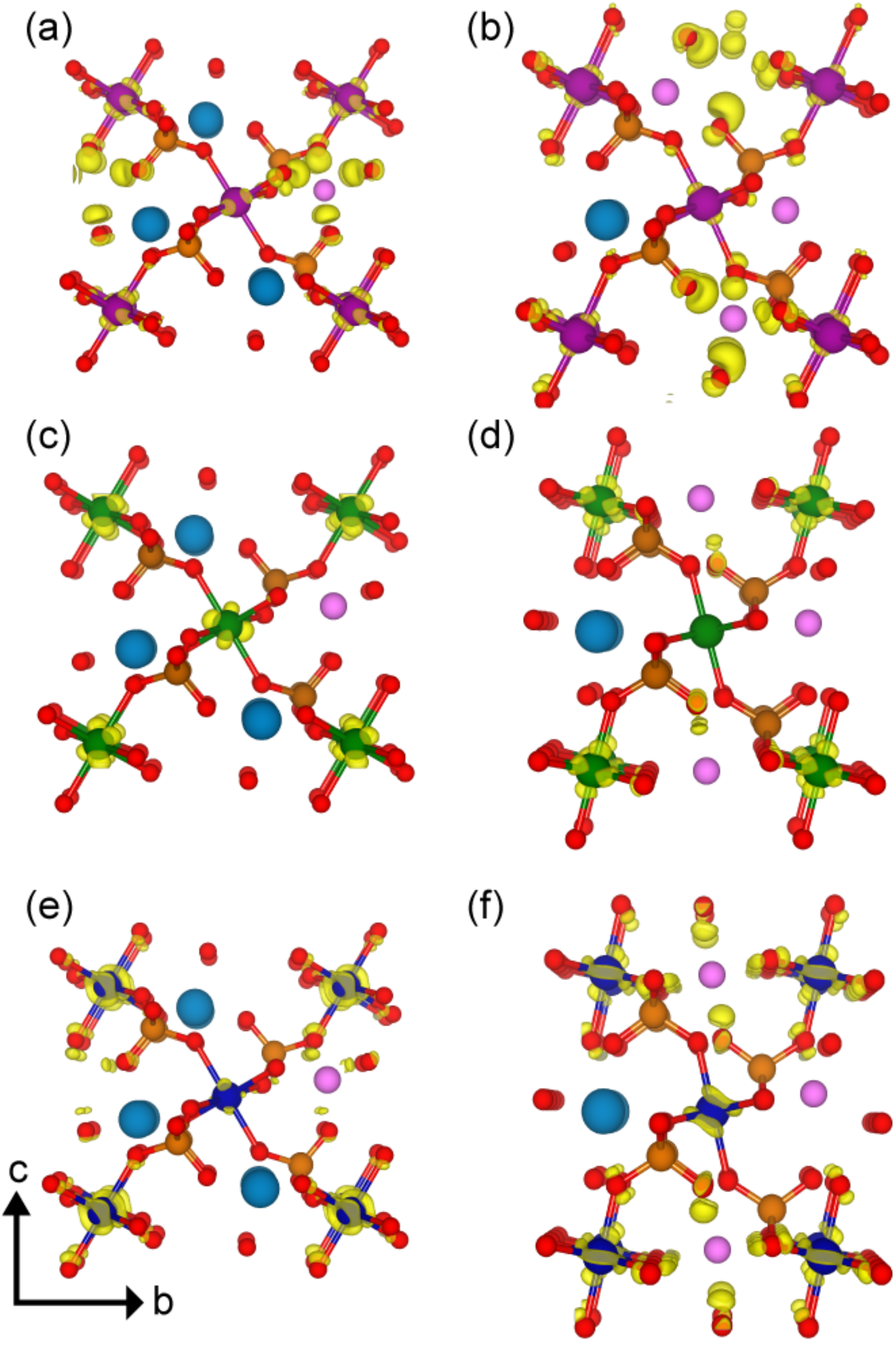

**Figure S8.** Electron density difference for Li-removal in (a) $Li_2Mn(SO_4)_2$, (b) $LiMn(SO_4)_2$, (c) $Li_2Fe(SO_4)_2$, (d) $LiFe(SO_4)_2$, (e) $Li_2Co(SO_4)_2$, and (f) $LiCo(SO_4)_2$. Red, blue, orange, purple, green, dark blue, and pink spheres represent O, Li, S, Mn, Fe, Co, and Li-vacancy, respectively. Isosurfaces are at 0.008 (Mn), 0.02 (Fe), and 0.015 (Co).

**Table S1**. Optimized lattice parameters of *o*-$Li_2Ni(SO_4)_2$ (M=Mn, Fe, Co, and Ni) estimated using SCAN+*U* and GGA+*U* approximation, compared with experimental values.

| System | | a (Å) | b (Å) | c (Å) | α (°) | β (°) | γ (°) | V(Å$^3$) |
|---|---|---|---|---|---|---|---|---|
| **$Li_2Mn(SO_4)_2$** | GGA+*U* | 9.51971 | 9.39573 | 14.03362 | 90.00 | 90.00 | 90.00 | 1255.23138 |
| | SCAN+*U* | 9.34368 | 9.24819 | 13.73874 | 90.00 | 90.00 | 90.00 | 1187.19290 |
| **$Li_2Fe(SO_4)_2$** | GGA+*U* | 9.38245 | 9.31127 | 13.86869 | 89.9999 | 90.0002 | 89.9999 | 1211.60352 |
| | SCAN+*U* | 9.18221 | 9.17010 | 13.61174 | 90.0001 | 90.0003 | 90.00 | 1146.13246 |
| | Expt. | 9.27980 | 9.20890 | 13.67650 | 90.00 | 90.00 | 90.00 | 1168.74938 |
| **$Li_2Co(SO_4)_2$** | GGA+*U* | 9.31610 | 9.20287 | 13.82582 | 90.00 | 90.00 | 90.00 | 118535436 |
| | SCAN+*U* | 9.16169 | 9.04557 | 13.57579 | 89.9996 | 90.00 | 89.998 | 1125.06175 |
| | Expt. | 9.20688 | 9.10175 | 13.71190 | 90.00 | 90.00 | 90.00 | 1149.03965 |
| **$Li_2Ni(SO_4)_2$** | GGA+*U* | 9.24625 | 9.13135 | 13.70520 | 90.00 | 90.00 | 90.00 | 1157.13982 |
| | SCAN+*U* | 9.09432 | 8.97466 | 13.45300 | 90.0001 | 90.00 | 90.00 | 1098.01195 |
| | Expt. | 9.12600 | 9.00100 | 13.54300 | 90.00 | 90.00 | 90.00 | 1112.46447 |

**Table S2**. Calculated and experimental average bond length, Baur's distortion index and bond angle variance of $SO_4$ and $MO_6$ polyhedra of *m*-$Li_2M(SO_4)_2$ (M=Mn, Fe, Co, and Ni).

| **System** | | **Bond-type** | **Average Bond Length (Å)** | **Distortion Index** | **Bond angle variance (°$^2$)** |
|---|---|---|---|---|---|
| **$Li_2Mn(SO_4)_2$** | GGA+*U* | Mn-O | 2.2124 | 0.02098 | 40.3915 |
| | | S-O | 1.4898 | 0.00408 | 3.2049 |
| | SCAN+*U* | Mn-O | 2.1750 | 0.01686 | 40.4581 |
| | | S-O | 1.4713 | 0.00420 | 3.1805 |
| | Expt. | Mn-O | 2.2029 | 0.01588 | 42.2134 |
| | | S-O | 1.4566 | 0.00543 | 4.3430 |
| **$Li_2Fe(SO_4)_2$** | GGA+*U* | Fe-O | 2.1625 | 0.01649 | 32.8736 |
| | | S-O | 1.4902 | 0.00525 | 4.2667 |
| | SCAN+*U* | Fe-O | 2.1290 | 0.01342 | 31.8448 |
| | | S-O | 1.4717 | 0.00529 | 3.9191 |
| | Expt. | Fe-O | 2.1036 | 0.01179 | 39.7825 |
| | | S-O | 1.4899 | 0.04329 | 28.2719 |
| **$Li_2Co(SO_4)_2$** | GGA+*U* | Co-O | 2.1351 | 0.01336 | 33.2996 |
| | | S-O | 1.4898 | 0.00624 | 4.0917 |
| | SCAN+*U* | Co-O | 2.1036 | 0.01024 | 33.5848 |
| | | S-O | 1.4710 | 0.00612 | 3.6361 |
| | Expt. | Co-O | 2.1100 | 0.00811 | 35.1295 |
| | | S-O | 1.4773 | 0.00776 | 3.1783 |
| **$Li_2Ni(SO_4)_2$** | GGA+*U* | Ni-O | 2.1099 | 0.01630 | 29.5633 |
| | | S-O | 1.4886 | 0.00659 | 4.2634 |
| | SCAN+*U* | Ni-O | 2.0738 | 0.01266 | 27.5136 |
| | | S-O | 1.4707 | 0.00686 | 3.7888 |
| | Expt. | Ni-O | 2.1776 | 0.02216 | 28.0094 |
| | | S-O | 1.4814 | 0.01109 | 46.9046 |

**Table S3.** Calculated and experimental average bond length, Baur's distortion index and bond angle variance of $SO_4$ and $MO_6$ polyhedra of *o*-$Li_2M(SO_4)_2$ (M=Mn, Fe, Co, and Ni).

| **System** | | **Bond-type** | **Average Bond Length (Å)** | **Distortion Index** | **Bond angle variance (°²)** |
|---|---|---|---|---|---|
| **$Li_2Mn(SO_4)_2$** | GGA+*U* | Mn-O | 2.2010 | 0.02046 | 20.2946 |
| | | S-O | 1.4914 | 0.00369 | 11.2927 |
| | SCAN+*U* | Mn-O | 2.1596 | 0.01774 | 20.7383 |
| | | S-O | 1.4742 | 0.00457 | 11.0910 |
| **$Li_2Fe(SO_4)_2$** | GGA+*U* | Fe-O | 2.1511 | 0.01364 | 20.2515 |
| | | S-O | 1.4912 | 0.00294 | 11.5389 |
| | SCAN+*U* | Fe-O | 2.1120 | 0.00973 | 23.3890 |
| | | S-O | 1.4746 | 0.00502 | 12.7614 |
| | Expt. | Fe-O | 2.1323 | 0.01657 | 24.4085 |
| | | S-O | 1.4929 | 0.00083 | 17.2093 |
| **$Li_2Co(SO_4)_2$** | GGA+*U* | Co-O | 2.1213 | 0.01813 | 18.03303 |
| | | S-O | 1.4932 | 0.00592 | 23.0551 |
| | SCAN+*U* | Co-O | 2.0858 | 0.01697 | 18.4334 |
| | | S-O | 1.4722 | 0.00226 | 12.2298 |
| | Expt. | Co-O | 2.0861 | 0.01286 | 17.1494 |
| | | S-O | 1.4831 | 0.01379 | 12.1005 |
| **$Li_2Ni(SO_4)_2$** | GGA+*U* | Ni-O | 2.0922 | 0.01518 | 15.3757 |
| | | S-O | 1.4927 | 0.00613 | 15.3757 |
| | SCAN+*U* | Ni-O | 2.0534 | 0.01051 | 15.8152 |
| | | S-O | 1.4745 | 0.00624 | 14.2162 |
| | Expt. | Ni-O | 2.0584 | 0.01314 | 15.6537 |
| | | S-O | 1.4736 | 0.00563 | 13.0296 |

**Table S4**. SCAN+*U* predicted on-site, averaged, magnetic moments of transition elements and corresponding electronic configurations.

| **System** | **Magnetic Moment** | **Corresponding Electronic configuration** |
|---|---|---|
| **$Li_2Mn(SO_4)_2$** | 4.657 | $t_{2g}^3e_g^2$ |
| **$LiMn(SO_4)_2$** | 3.889 | $t_{2g}^3e_g^1$ |
| **$Mn(SO_4)_2$** | 3.033 | $t_{2g}^3e_g^0$ |
| **$Li_2Fe(SO_4)_2$** | 3.745 | $t_{2g}^4e_g^2$ |
| **$LiFe(SO_4)_2$** | 4.361 | $t_{2g}^3e_g^2$ |
| **$Fe(SO_4)_2$** | 4.301 | $t_{2g}^3e_g^2$ |
| **$Li_2Co(SO_4)_2$** | 2.802 | $t_{2g}^5e_g^2$ |
| **$LiCo(SO_4)_2$** | 3.224 | $t_{2g}^4e_g^2$ |
| **$Co(SO_4)_2$** | 3.236 | $t_{2g}^4e_g^2$ |
| **$Li_2Ni(SO_4)_2$** | 1.785 | $t_{2g}^6e_g^2$ |
| **$LiNi(SO_4)_2$** | 1.857 | $t_{2g}^6e_g^2$ |
| **$Ni(SO_4)_2$** | 1.831 | $t_{2g}^6e_g^2$ |

**Table S5.** SCAN+*U* predicted change in on-site magnetic moment of individual oxygen ions at various (de)lithiation states of $Li_xNi(SO_4)_2$.

| | $Li_2Ni(SO_4)_2$ | $LiNi(SO_4)_2$ | $Ni(SO_4)_2$ |
|---|---|---|---|
| O1 | -0.001 | 0.053 | **0.520** |
| O2 | -0.001 | **0.324** | **0.519** |
| O3 | -0.001 | 0.053 | **0.520** |
| O4 | -0.001 | **0.323** | **0.519** |
| O5 | 0.030 | 0.051 | 0.073 |
| O6 | 0.029 | 0.078 | 0.075 |
| O7 | 0.029 | 0.051 | 0.073 |
| O8 | 0.030 | 0.078 | 0.074 |
| O9 | 0.026 | 0.038 | 0.060 |
| O10 | 0.025 | 0.061 | 0.059 |
| O11 | 0.025 | 0.038 | 0.060 |
| O12 | 0.026 | 0.061 | 0.060 |
| O13 | 0.029 | 0.064 | **0.314** |
| O14 | 0.029 | **0.349** | **0.314** |
| O15 | 0.029 | 0.063 | **0.314** |
| O16 | 0.029 | **0.349** | **0.314** |

**Table S6.** SCAN+*U* predicted change in on-site magnetic moment of individual oxygen ions at various (de)lithiation states of $Li_xMn(SO_4)_2$.

| | $Li_2Mn(SO_4)_2$ | $LiMn(SO_4)_2$ | $Mn(SO_4)_2$ |
|---|---|---|---|
| O1 | 0 | 0.001 | 0.009 |
| O2 | 0 | 0.001 | 0.008 |
| O3 | 0 | 0.001 | 0.009 |
| O4 | 0 | 0.001 | 0.008 |
| O5 | 0.018 | 0.011 | -0.023 |
| O6 | 0.018 | 0.015 | -0.025 |
| O7 | 0.018 | 0.011 | -0.023 |
| O8 | 0.018 | 0.015 | -0.024 |
| O9 | 0.017 | -0.003 | -0.015 |
| O10 | 0.017 | -0.005 | -0.016 |
| O11 | 0.017 | -0.003 | -0.015 |
| O12 | 0.017 | -0.005 | -0.016 |
| O13 | 0.017 | -0.021 | -0.024 |
| O14 | 0.017 | -0.035 | -0.026 |
| O15 | 0.017 | -0.021 | -0.024 |
| O16 | 0.017 | -0.034 | -0.025 |

**Table S7.** SCAN+*U* predicted change in on-site magnetic moment of individual oxygen ions at various (de)lithiation states of $Li_xCo(SO_4)_2$.

| | $Li_2Co(SO_4)_2$ | $LiCo(SO_4)_2$ | $Co(SO_4)_2$ |
|---|---|---|---|
| O1 | 0 | 0.009 | 0.276 |
| O2 | 0 | 0.020 | 0.22 |
| O3 | 0 | 0.009 | 0.22 |
| O4 | 0 | 0.020 | 0.276 |
| O5 | 0.025 | 0.081 | 0.154 |
| O6 | 0.025 | 0.102 | 0.176 |
| O7 | 0.025 | 0.081 | 0.176 |
| O8 | 0.025 | 0.102 | 0.154 |
| O9 | 0.020 | 0.068 | 0.185 |
| O10 | 0.019 | 0.11 | 0.176 |
| O11 | 0.019 | 0.068 | 0.176 |
| O12 | 0.020 | 0.11 | 0.185 |
| O13 | 0.023 | 0.099 | 0.192 |
| O14 | 0.023 | 0.184 | 0.199 |
| O15 | 0.023 | 0.099 | 0.199 |
| O16 | 0.023 | 0.184 | 0.192 |

**Table S8.** SCAN+*U* predicted change in on-site magnetic moment of individual oxygen ions at various (de)lithiation states of $Li_xFe(SO_4)_2$.

| | $Li_2Fe(SO_4)_2$ | $LiFe(SO_4)_2$ | $Fe(SO_4)_2$ |
|---|---|---|---|
| O1 | 0 | 0.006 | -0.180 |
| O2 | 0 | 0.006 | -0.185 |
| O3 | 0 | 0.006 | -0.180 |
| O4 | 0 | 0.007 | -0.184 |
| O5 | 0.021 | 0.086 | 0.006 |
| O6 | 0.021 | 0.085 | 0.003 |
| O7 | 0.021 | 0.085 | 0.006 |
| O8 | 0.021 | 0.086 | 0.002 |
| O9 | 0.020 | 0.078 | 0.003 |
| O10 | 0.020 | 0.078 | -0.004 |
| O11 | 0.020 | 0.078 | 0.003 |
| O12 | 0.020 | 0.079 | -0.004 |
| O13 | 0.024 | 0.075 | -0.005 |
| O14 | 0.024 | 0.081 | -0.003 |
| O15 | 0.024 | 0.076 | -0.005 |
| O16 | 0.024 | 0.081 | -0.003 |

**Table S9**. (a) Mulliken and (b) Löwdin charge populations at various (de)lithiation states of *m*-$Li_xNi(SO_4)_2$, calculated using LOBSTER package, based on SCAN+*U*-computed charge densities.

**(a)**

| Element | $Li_2Ni(SO_4)_2$ | $LiNi(SO_4)_2$ | $Ni(SO_4)_2$ |
|---|---|---|---|
| Ni1 | 1.51 | 1.58 | 1.60 |
| Ni2 | 1.51 | 1.58 | 1.60 |
| O1 | -1.14 | -1.06 | -0.67 |
| O2 | -1.14 | -0.87 | -0.67 |
| O3 | -1.14 | -1.06 | -0.67 |
| O4 | -1.14 | -0.87 | -0.67 |
| O5 | -1.08 | -1.07 | -1.03 |
| O6 | -1.08 | -1.02 | -1.03 |
| O7 | -1.08 | -1.07 | -1.03 |
| O8 | -1.08 | -1.02 | -1.03 |
| O9 | -1.11 | -1.11 | -1.06 |
| O10 | -1.11 | -1.05 | -1.06 |
| O11 | -1.11 | -1.11 | -1.06 |
| O12 | -1.11 | -1.05 | -1.06 |
| O13 | -1.10 | -1.09 | -0.91 |
| O14 | -1.10 | -0.86 | -0.90 |
| O15 | -1.10 | -1.09 | -0.91 |
| O16 | -1.10 | -0.86 | -0.90 |

**(b)**

| Element | $Li_2Ni(SO_4)_2$ | $LiNi(SO_4)_2$ | $Ni(SO_4)_2$ |
|---|---|---|---|
| Ni1 | 1.39 | 1.44 | 1.42 |
| Ni2 | 1.39 | 1.44 | 1.42 |
| O1 | -1.03 | -0.97 | -0.62 |
| O2 | -1.03 | -0.78 | -0.62 |
| O3 | -1.03 | -0.97 | -0.62 |
| O4 | -1.03 | -0.78 | -0.62 |
| O5 | -1.00 | -0.99 | -0.94 |
| O6 | -1.00 | -0.93 | -0.95 |
| O7 | -1.00 | -0.99 | -0.94 |
| O8 | -1.00 | -0.93 | -0.95 |
| O9 | -1.02 | -1.02 | -0.97 |
| O10 | -1.02 | -0.96 | -0.97 |
| O11 | -1.02 | -1.02 | -0.97 |
| O12 | -1.02 | -0.96 | -0.97 |
| O13 | -1.01 | -1.00 | -0.83 |
| O14 | -1.01 | -0.78 | -0.83 |
| O15 | -1.01 | -1.00 | -0.83 |
| O16 | -1.01 | -0.78 | -0.83 |

**Table S10**. Mulliken charge population at various (de)lithiation states of *m*-$Li_xMn(SO_4)_2$ calculated using LOBSTER package, based on SCAN+*U*-computed charge densities.

| Element | $Li_2Mn(SO_4)_2$ | $LiMn(SO_4)_2$ | $Mn(SO_4)_2$ |
|---|---|---|---|
| Mn1 | 1.61 | 1.92 | 2.05 |
| Mn2 | 1.61 | 1.92 | 2.05 |
| O1 | -1.14 | -1.07 | -0.97 |
| O2 | -1.14 | -1.07 | -0.97 |
| O3 | -1.14 | -1.07 | -0.97 |
| O4 | -1.14 | -1.07 | -0.97 |
| O5 | -1.10 | -1.06 | -0.95 |
| O6 | -1.10 | -1.05 | -0.94 |
| O7 | -1.10 | -1.06 | -0.95 |
| O8 | -1.10 | -1.05 | -0.94 |
| O9 | -1.11 | -1.08 | -0.96 |
| O10 | -1.11 | -1.03 | -0.97 |
| O11 | -1.11 | -1.08 | -0.96 |
| O12 | -1.11 | -1.03 | -0.97 |
| O13 | -1.10 | -1.05 | -0.94 |
| O14 | -1.10 | -0.99 | -0.94 |
| O15 | -1.10 | -0.105 | -0.94 |
| O16 | -1.10 | -0.99 | -0.94 |

**Table S11**. Mulliken charge population at various (de)lithiation states states of *m*-$Li_xFe(SO_4)_2$ calculated using LOBSTER package, based on SCAN+*U*-computed charge densities.

| Element | $Li_2Fe(SO_4)_2$ | $LiFe(SO_4)_2$ | $Fe(SO_4)_2$ |
|---|---|---|---|
| Fe1 | 1.58 | 2.09 | 2.14 |
| Fe2 | 1.58 | 2.09 | 2.14 |
| O1 | -1.14 | -1.08 | -0.89 |
| O2 | -1.14 | -1.08 | -0.88 |
| O3 | -1.14 | -1.08 | -0.89 |
| O4 | -1.14 | -1.08 | -0.88 |
| O5 | -1.09 | -1.06 | -1.01 |
| O6 | -1.09 | -1.06 | -1.01 |
| O7 | -1.09 | -1.06 | -1.01 |
| O8 | -1.09 | -1.06 | -1.01 |
| O9 | -1.11 | -1.08 | -1.02 |
| O10 | -1.11 | -1.07 | -1.02 |
| O11 | -1.11 | -1.08 | -1.02 |
| O12 | -1.11 | -1.07 | -1.02 |
| O13 | -1.11 | -1.07 | -1.01 |
| O14 | -1.11 | -1.08 | -1.01 |
| O15 | -1.11 | -1.07 | -1.01 |
| O16 | -1.11 | -1.08 | -1.01 |

**Table S12**. Mulliken charge population at various (de)lithiation states of *m*-$Li_2Co(SO_4)_2$ calculated using LOBSTER package, based on SCAN+*U*-computed charge densities.

| Element | $Li_2Co(SO_4)_2$ | $LiCo(SO_4)_2$ | $Co(SO_4)_2$ |
|---|---|---|---|
| Co1 | 1.57 | 1.90 | 1.97 |
| Co2 | 1.57 | 1.90 | 1.97 |
| O1 | -1.14 | -1.07 | -0.76 |
| O2 | -1.14 | -1.07 | -0.80 |
| O3 | -1.14 | -1.07 | -0.80 |
| O4 | -1.14 | -1.07 | -0.76 |
| O5 | -1.09 | -1.02 | -0.92 |
| O6 | -1.09 | -1.02 | -0.91 |
| O7 | -1.09 | -1.02 | -0.91 |
| O8 | -1.09 | -1.02 | -0.92 |
| O9 | -1.11 | -1.05 | -0.92 |
| O10 | -1.11 | -1.05 | -0.93 |
| O11 | -1.11 | -1.05 | -0.93 |
| O12 | -1.11 | -1.05 | -0.92 |
| O13 | -1.11 | -1.05 | -0.90 |
| O14 | -1.11 | -1.05 | -0.90 |
| O15 | -1.11 | -1.05 | -0.90 |
| O16 | -1.11 | -1.05 | -0.90 |

**Table S13**. SCAN+*U*-calculated S-O and M-O bond lengths at various (de)lithiation states for *m*-$Li_xM(SO_4)_2$ (M = Ni, Mn, Fe and Co).

| $Li_2Ni(SO_4)_2$ | $LiNi(SO_4)_2$ | $Ni(SO_4)_2$ |
|---|---|---|
| (Ni1-O6) = 2.12365 Å | (Ni1-O6) = 2.11576 Å | (Ni1-O6) = 2.03590 Å |
| (Ni1-O11) = 2.03924 Å | (Ni1-O11) = 2.05477 Å | (Ni1-O11) = 2.03518 Å |
| (Ni1-O16) = 2.06790 Å | (Ni1-O16) = 1.99274 Å | (Ni1-O16) = 2.02626 Å |
| (Ni1-O15) = 2.07702 Å | (Ni1-O15) = 2.04683 Å | (Ni1-O15) = 2.03603 Å |
| (Ni1-O12) = 2.03538 Å | (Ni1-O12) = 2.03403 Å | (Ni1-O12) = 2.03341 Å |
| (Ni1-O5) = 2.09935 Å | (Ni1-O5) = 2.04563 Å | (Ni1-O5) = 2.03062 Å |
| | | |
| (Ni2-O13) = 2.06790 Å | (Ni2-O13) = 2.04667 Å | (Ni2-O13) = 2.03546 Å |
| (Ni2-O7) = 2.12364 Å | (Ni2-O7) = 2.04545 Å | (Ni2-O7) = 2.03094 Å |
| (Ni2-O10) = 2.03922 Å | (Ni2-O10) = 2.03440 Å | (Ni2-O10) = 2.03375 Å |
| (Ni2-O9) = 2.03538 Å | (Ni2-O9) = 2.05459 Å | (Ni2-O9) = 2.03502 Å |
| (Ni2-O8) = 2.09936 Å | (Ni2-O8) = 2.11490 Å | (Ni2-O8) = 2.03557 Å |
| (Ni2-O14) = 2.07705 Å | (Ni2-O14) = 1.99412 Å | (Ni2-O14) = 2.02644 Å |

| $Li_2Ni(SO_4)_2$ | $LiNi(SO_4)_2$ | $Ni(SO_4)_2$ |
|---|---|---|
| (S1-O5) = 1.48130(0) Å | (S1-O5) = 1.48377 Å | (S1-O5) = 1.44933 Å |
| (S1-O1) = 1.44990(0) Å | (S1-O1) = 1.43421 Å | (S1-O1) = 1.49229 Å |
| (S1-O9) = 1.47531(0) Å | (S1-O9) = 1.48185 Å | (S1-O9) = 1.44386 Å |
| (S1-O13) = 1.47506(0) Å | (S1-O13)= 1.48163 Å | (S1-O13) = 1.48141 Å |
| | | |
| (S2-O14) = 1.47518(0) Å | (S2-O14) = 1.50160 Å | (S2-O14)= 1.48176 Å |
| (S2-O10) = 1.47631(0) Å | (S2-O10) = 1.44639 Å | (S2-O10)= 1.44401 Å |
| (S2-O2) = 1.45054(0) Å | (S2-O2) = 1.47039 Å | (S2-O2) = 1.49205 Å |
| (S2-O6) = 1.48086(0) Å | (S2-O6) = 1.45015 Å | (S2-O6) = 1.44891 Å |
| | | |
| (S3-O15) = 1.47518(0) Å | (S3-O15) = 1.48162 Å | (S3-O15)= 1.48140 Å |
| (S3-O7) = 1.48086(0) Å | (S3-O7) = 1.48372 Å | (S3-O7) = 1.44932 Å |
| (S3-O11) = 1.47630(0) Å | (S3-O11) = 1.48187 Å | (S3-O11)= 1.44394 Å |
| (S3-O3) = 1.45054(0) Å | (S3-O3) = 1.43413 Å | (S3-O3) = 1.49232 Å |
| | | |
| (S4-O4) = 1.44992(0) Å | (S4-O4) = 1.47011 Å | (S4-O4) = 1.49202 Å |
| (S4-O12) = 1.47530(0) Å | (S4-O12) = 1.44643 Å | (S4-O12) = 1.44413 Å |
| (S4-O8) = 1.48129(0) Å | (S4-O8) = 1.45027 Å | (S4-O8) = 1.44879 Å |
| (S4-O16) = 1.47503(0) Å | (S4-O16) = 1.50196 Å | (S4-O16) = 1.48170 Å |

| $Li_2Mn(SO_4)_2$ | $LiMn(SO_4)_2$ | $Mn(SO_4)_2$ |
|---|---|---|
| (Mn1-O6) = 2.18736 Å | (Mn1-O6) = 2.17687 Å | (Mn1-O6) = 1.91301 Å |
| (Mn1-O11) = 2.12220 Å | (Mn1-O11) = 1.97186 Å | (Mn1-O11) = 1.90916 Å |
| (Mn1-O16) = 2.22025 Å | (Mn1-O16) = 1.90674 Å | (Mn1-O16) = 1.90003 Å |
| (Mn1-O15) = 2.22014 Å | (Mn1-O15) = 1.98723 Å | (Mn1-O15) = 1.90020 Å |
| (Mn1-O12) = 2.12210 Å | (Mn1-O12) = 1.92036 Å | (Mn1-O12) = 1.91823 Å |
| (Mn1-O5) = 2.18763 Å | (Mn1-O5) = 2.30486 Å | (Mn1-O5) = 1.91517 Å |
| | | |
| (Mn2-O13) = 2.22025 Å | (Mn2-O13) = 1.98704 Å | (Mn2-O13) = 1.89995 Å |
| (Mn2-O7) = 2.18736 Å | (Mn2-O7) = 2.30458 Å | (Mn2-O7) = 1.91544 Å |
| (Mn2-O10) = 2.12220 Å | (Mn2-O10) = 1.92135 Å | (Mn2-O10) = 1.91799 Å |
| (Mn2-O9) = 2.12210 Å | (Mn2-O9) = 1.97243 Å | (Mn2-O9) = 1.90963 Å |
| (Mn2-O8) = 2.18765 Å | (Mn2-O8) = 2.17668 Å | (Mn2-O8) = 1.91274 Å |
| (Mn2-O14) = 2.22014 Å | (Mn2-O14) = 1.90618 Å | (Mn2-O14) = 1.90043 Å |

| $Li_2Mn(SO_4)_2$ | $LiMn(SO_4)_2$ | $Mn(SO_4)_2$ |
|---|---|---|
| (S1-O5) = 1.48012 Å | (S1-O5) = 1.46125 Å | (S1-O5) = 1.49804 Å |
| (S1-O1) = 1.45904 Å | (S1-O1) = 1.41929 Å | (S1-O1) = 1.40121 Å |
| (S1-O9) = 1.47288 Å<br>(S1-O13) = 1.47374 Å | (S1-O9) = 1.50712 Å<br>(S1-O13) = 1.50285 Å | (S1-O9) = 1.50762 Å<br>(S1-O13) = 1.50113 Å |
| (S2-O14) = 1.47375 Å<br>(S2-O10) = 1.47285 Å<br>(S2-O2) = 1.45904 Å<br>(S2-O6) = 1.48013 Å | (S2-O14) = 1.50770 Å<br>(S2-O10) = 1.49699 Å<br>(S2-O2) = 1.43209 Å<br>(S2-O6) = 1.45237 Å | (S2-O14) = 1.50201 Å<br>(S2-O10) = 1.50613 Å<br>(S2-O2) = 1.40126 Å<br>(S2-O6) = 1.49914 Å |
| (S3-O15) = 1.47376 Å<br>(S3-O7) = 1.48015 Å<br>(S3-O11) = 1.47287 Å<br>(S3-O3) = 1.45897 Å | (S3-O15) = 1.50279 Å<br>(S3-O7) = 1.46124 Å<br>(S3-O3) = 1.41928 Å<br>(S3-O11) = 1.50721 Å | (S3-O15) = 1.50137 Å<br>(S3-O7) = 1.49814 Å<br>(S3-O11) = 1.50798 Å<br>(S3-O3) = 1.40121 Å |
| (S4-O4) = 1.45899 Å<br>(S4-O12) = 1.47290 Å<br>(S4-O8) = 1.48013 Å<br>(S4-O16) = 1.47375 Å | (S4-O4) = 1.43213 Å<br>(S4-O12) = 1.49714 Å<br>(S4-O8) = 1.45236 Å<br>(S4-O16) = 1.50749 Å | (S4-O4) = 1.40130 Å<br>(S4-O12) = 1.50599 Å<br>(S4-O8) = 1.49912 Å<br>(S4-O16) = 1.50203 Å |

| $Li_2Fe(SO_4)_2$ | $LiFe(SO_4)_2$ | $Fe(SO_4)_2$ |
|---|---|---|
| (Fe1-O6) = 2.17040 Å<br>(Fe1-O11) = 2.10094 Å<br>(Fe1-O16) = 2.11517 Å<br>(Fe1-O15) = 2.11858 Å<br>(Fe1-O12) = 2.09559 Å<br>(Fe1-O5) = 2.17334 Å | (Fe1-O6) = 1.99780 Å<br>l(Fe1-O11) = 2.03597 Å<br>l(Fe1-O16) = 2.02310 Å<br>l(Fe1-O15) = 2.04369 Å<br>l(Fe1-O12) = 2.03401 Å<br>l(Fe1-O5) = 1.98894 Å | (Fe1-O6) = 1.98399 Å<br>(Fe1-O11) = 1.98263 Å<br>(Fe1-O16) = 1.98201 Å<br>(Fe1-O15) = 1.98580 Å<br>(Fe1-O12) = 1.99253 Å<br>(Fe1-O5) = 1.97956 Å |
| (Fe2-O13) = 2.11517 Å<br>(Fe2-O7) = 2.17054 Å<br>(Fe2-O10) = 2.10104 Å<br>(Fe2-O9) = 2.09559 Å<br>(Fe2-O8) = 2.17354 Å<br>(Fe2-O14) = 2.11824 Å | (Fe2-O13) = 2.04524 Å<br>(Fe2-O7) = 1.99311 Å<br>(Fe2-O10) = 2.03242 Å<br>(Fe2-O9) = 2.03567 Å<br>(Fe2-O8) = 1.99551 Å<br>(Fe2-O14) = 2.02075 Å | (Fe2-O13) = 1.98595 Å<br>(Fe2-O7) = 1.97946 Å<br>(Fe2-O10) = 1.99247 Å<br>(Fe2-O9) = 1.98282 Å<br>(Fe2-O8) = 1.98407 Å<br>(Fe2-O14) = 1.98197 Å |

| $Li_2Fe(SO_4)_2$ | $LiFe(SO_4)_2$ | $Fe(SO_4)_2$ |
|---|---|---|
| (S1-O5) = 1.47663 Å<br>(S1-O1) = 1.45590 Å<br>(S1-O9) = 1.47299 Å<br>(S1-O13) = 1.48077 Å | (S1-O5) = 1.47288 Å<br>(S1-O9) = 1.49002 Å<br>(S1-O1) = 1.42655 Å<br>(S1-O13) = 1.49631 Å | l(S1-O5) = 1.47878 Å<br>l(S1-O9) = 1.48012 Å<br>l(S1-O1) = 1.43542 Å<br>l(S1-O13) = 1.47920 Å |
| (S2-O14) = 1.48077 Å<br>(S2-O10) = 1.47266 Å<br>(S2-O2) = 1.45608 Å<br>(S2-O6) = 1.47713 Å | (S2-O14) = 1.49091 Å<br>(S2-O2) = 1.42697 Å<br>(S2-O10) = 1.49730 Å<br>(S2-O6) = 1.47106 Å | (S2-O14) = 1.48023 Å<br>(S2-O2) = 1.43606 Å<br>(S2-O10) = 1.47905 Å<br>(S2-O6) = 1.47835 Å |
| (S3-O15) = 1.48071 Å<br>(S3-O7) = 1.47711 Å<br>(S3-O11) = 1.47266 Å<br>(S3-O3) = 1.45613 Å | (S3-O15) = 1.49504 Å<br>(S3-O7) = 1.47206 Å<br>(S3-O11) = 1.49226 Å<br>(S3-O3) = 1.42691 Å | (S3-O15) = 1.47921 Å<br>(S3-O7) = 1.47865 Å<br>(S3-O11) = 1.48008 Å<br>(S3-O3) = 1.43557 Å |
| (S4-O4) = 1.45597 Å<br>(S4-O12) = 1.47296 Å<br>(S4-O8) = 1.47658 Å<br>(S4-O16) = 1.48077 Å | (S4-O4) = 1.42707 Å<br>(S4-O12) = 1.49524 Å<br>(S4-O8) = 1.47191 Å<br>(S4-O16) = 1.49293 Å | (S4-O4) = 1.43588 Å<br>(S4-O12) = 1.47889 Å<br>(S4-O8) = 1.47850 Å<br>(S4-O16) = 1.48026 Å |

| $Li_2Co(SO_4)_2$ | $LiCo(SO_4)_2$ | $Co(SO_4)_2$ |
|---|---|---|
| (Co1-O6) = 2.12335 Å | (Co1-O6) = 2.05906 Å | (Co1-O6) = 2.00399 Å |
| (Co1-O11) = 2.07338 Å<br>(Co1-O16) = 2.11503 Å<br>(Co1-O15) = 2.11641 Å<br>(Co1-O12) = 2.06921 Å<br>(Co1-O5) = 2.12421 Å<br><br>(Co2-O13) = 2.11504 Å<br>(Co2-O7) = 2.12335 Å<br>(Co2-O10) = 2.07339 Å<br>(Co2-O9) = 2.06921 Å<br>(Co2-O8) = 2.12420 Å<br>(Co2-O14) = 2.11642 Å | (Co1-O11) = 2.06748 Å<br>(Co1-O16) = 1.87982 Å<br>(Co1-O15) = 1.99580 Å<br>(Co1-O12) = 1.95123 Å<br>(Co1-O5) = 2.08089 Å<br><br>(Co2-O13) = 1.99526 Å<br>(Co2-O7) = 2.08059 Å<br>(Co2-O10) = 1.95141 Å<br>(Co2-O9) = 2.06788 Å<br>(Co2-O8) = 2.05876 Å<br>(Co2-O14) = 1.87980 Å | (Co1-O11) = 1.99577 Å<br>(Co1-O16) = 1.92431 Å<br>(Co1-O15) = 1.90834 Å<br>(Co1-O12) = 2.00657 Å<br>(Co1-O5) = 2.00595 Å<br><br>(Co2-O13) = 1.92429 Å<br>(Co2-O7) = 2.00396 Å<br>(Co2-O10) = 1.99584 Å<br>(Co2-O9) = 2.00661 Å<br>(Co2-O8) = 2.00599 Å<br>(Co2-O14) = 1.90832 Å |

| $Li_2Co(SO_4)_2$ | $LiCo(SO_4)_2$ | $Co(SO_4)_2$ |
|---|---|---|
| (S1-O5) = 1.48201 Å<br>(S1-O1) = 1.45300 Å<br>(S1-O9) = 1.47362 Å<br>(S1-O13) = 1.47535 Å<br><br>(S2-O14) = 1.47538 Å<br>(S2-O10) = 1.47350 Å<br>(S2-O2) = 1.45310 Å<br>(S2-O6) = 1.48221 Å<br><br>(S3-O15) = 1.47538 Å<br>(S3-O7) = 1.48221 Å<br>(S3-O11) = 1.47350 Å<br>(S3-O3) = 1.45310 Å<br><br>(S4-O4) = 1.45301 Å<br>(S4-O12) = 1.47361 Å<br>(S4-O8) = 1.48202 Å<br>(S4-O16) = 1.47535 Å | (S1-O5) = 1.47676(0) Å<br>(S1-O1) = 1.41797(0) Å<br>(S1-O9) = 1.49453(0) Å<br>(S1-O13) = 1.49647(0) Å<br><br>(S2-O14) = 1.50678(0) Å<br>(S2-O2) = 1.43067(0) Å<br>(S2-O6) = 1.46841(0) Å<br>(S2-O10) = 1.47793(0) Å<br><br>(S3-O15) = 1.49652(0) Å<br>(S3-O7) = 1.47681(0) Å<br>(S3-O11) = 1.49461(0) Å<br>(S3-O3) = 1.41794(0) Å<br><br>(S4-O4) = 1.43063(0) Å<br>(S4-O12) = 1.47802(0) Å<br>(S4-O8) = 1.46844(0) Å<br>(S4-O16) = 1.50675(0) Å | (S1-O5) = 1.46576 Å<br>(S1-O9) = 1.47448 Å<br>(S1-O1) = 1.44833 Å<br>(S1-O13) = 1.48578 Å<br><br>(S2-O14) = 1.49110 Å<br>(S2-O2) = 1.43831 Å<br>(S2-O10) = 1.47569 Å<br>(S2-O6) = 1.47211 Å<br><br>(S3-O15) = 1.49121 Å<br>(S3-O7) = 1.47211 Å<br>(S3-O11) = 1.47585 Å<br>(S3-O3) = 1.43838 Å<br><br>(S4-O4) = 1.44849 Å<br>(S4-O12) = 1.47458 Å<br>(S4-O8) = 1.46577 Å<br>(S4-O16) = 1.48560 Å |